\documentclass[pra, twocolumn, showpacs, nofootinbibfloatfix, amsmath, amsfonts, amssymb, table, xcdraw]{revtex4-1}%
\usepackage[utf8]{inputenc}
\usepackage{amsmath,amsfonts,amssymb,color}
\usepackage{amsthm}
\usepackage{caption}
\usepackage{subcaption}
\usepackage{leftidx}
\usepackage{caption}
\usepackage{graphicx}
\usepackage{xcolor}
\usepackage{dcolumn}
\usepackage{comment}
\usepackage{bm}
\usepackage{epstopdf}
\usepackage{epsfig}
\usepackage{mathdots} 
\usepackage{subfig}
\usepackage{verbatim}
\def\CH{\textcolor{red}}
\usepackage{comment}
\usepackage{verbatim}
\newcommand{\dd}[1]{\mathrm{d}#1}

\NewDocumentCommand{\evalat}{sO{\big}mm}{%
  \IfBooleanTF{#1}
   {\mleft. #3 \mright|_{#4}}
   {#3#2|_{#4}}%
}

\usepackage{caption}
\usepackage{graphicx}
\usepackage{subcaption}
\usepackage{braket}
\usepackage{amssymb}
\usepackage{mathrsfs}
\usepackage{cleveref}
\creflabelformat{equation}{(#2#1#3)}
\crefrangeformat{equation}{(#3#1#4--#5#2#6)}
\usepackage{amssymb}

\usepackage{chngcntr}
\usepackage{float}

\usepackage{lipsum} % for dummy text
\usepackage{subcaption}
\usepackage{placeins} % For \FloatBarrier
\usepackage{tikz}  % For drawing
\usepackage{array} % For advanced table features

\usepackage{xcolor}
\usepackage{pifont} % For additional symbols
\usepackage{multirow}  % For multirow functionality
\usepackage{array}     % For advanced table features
\usepackage{colortbl}  % For coloring cells

\usepackage{titlesec}
\usepackage{lipsum} % for dummy text

\begin{document}
	\title{The effect of staggered nonlinearity on the Su-Schrieffer-Heeger model}
        \author{Ahmed Alharthy}
        \email{Ahmedal7@protonmail.com}
        \affiliation{%
		Department of Physics, King Fahd University of Petroleum and Minerals, 31261 Dhahran, Saudi Arabia
	}
	\author{Raditya Weda Bomantara}
	\email{Corresponding Author: Raditya.Bomantara@kfupm.edu.sa}
	\affiliation{%
		Department of Physics, Interdisciplinary Research Center for Advanced Quantum Computing, King Fahd University of Petroleum and Minerals, 31261 Dhahran, Saudi Arabia
	}
	\date{\today}
	
	%%%%%%%%%%%%%%%%%%%% ABSTRACT %%%%%%%%%%%%%%%%%%%%%%%%
	%\begin{linenumbers}
	
	\vspace{2cm}
	
\begin{abstract}
We investigate the spectral properties of the Su-Schrieffer-Heeger (SSH) model with sublattice-dependent onsite nonlinearity. Two complementary approaches are employed in our studies. First, Bloch state solutions under periodic boundary conditions are assumed to enable semi-analytical treatment, which allows us to obtain the system's energy band structure and further derive a general expression of the Zak phase that incorporates nonlinearity-induced correction (referred to as nonlinear Zak phase). This analysis reveals that, at sufficiently high nonlinearities, a nonlinearity-induced topological phase transition occurs, marked by a discontinuity in the nonlinear Zak phase. The second approach amounts to numerically obtaining other (non-Bloch) solutions under open boundary conditions, employing the Self-Consistent Field Iterative Method. Its main results include the observation of an edge state's energy that is independent of a nonlinear parameter, a persisting band touching point that only shifts in the presence of perturbations reminiscent of Weyl points in a Weyl semimetal, as well as delocalized solutions that persist even at extreme nonlinearity strengths. These findings illuminate the rich interplay between topology and nonlinearity in lattice models with potential realization in optical/acoustic waveguide settings.
\end{abstract}

\maketitle

\section{Introduction} 
\label{intro}

Topological systems represent paradigmatic phases of matter in which an otherwise abstract mathematical concept of topology manifests itself as some physical observables \cite{Exp. Realization FQH, Vacuum degeneracy, Topological Orders in Rigid States, Superconductors are topologically ordered, Quantum Spin Hall Insulator State in HgTe Quantum Wells}. Of particular example is the so-called topological insulator \cite{Quantum Spin Hall Insulator State in HgTe Quantum Wells}, characterized by the presence of robust in-gap states at its boundaries, termed the edge states, that have a topological origin. In two and three dimensions, these edge states are typically propagating in nature, thereby holding promise for potential use as dissipation-free and consequently energy efficient electronic/spintronic devices.

The Su–Schrieffer–Heeger (SSH) model \cite{SSH} is a one dimensional model that is often used as a textbook example of topological phase. It describes a tight-binding system of particles hopping on a lattice with alternating hopping amplitudes. Depending on the ratio between these hopping amplitudes, the system can be classified as either topologically nontrivial, which supports a pair of zero energy edge states, or topologically trivial, which does not support edge states. Apart from the existence of edge states, the topology of the SSH model can be completely captured via the Zak phase \cite{ZakPhase} of its energy eigenstates, which is quantized to either zero (topologically trivial phase) or $\pi$ (topologically nontrivial phase). In addition to its mathematical simplicity, the SSH model is also easy to be experimentally realize. Indeed, the SSH model has been realized experimentally in various platforms including Rydberg atoms \cite{Rydberg atoms}, \newline 
 \newline \newline in addition to cold-atomic \cite{AtomicRealization}, photonic \cite{PhotonicExpRealization}, electric circuit \cite{ElectricCircuit Realization}, superconducting circuit \cite{superconductingRealization1, superconductingRealization2}, acoustic \cite{Acoustic}, as well as mechanical systems \cite{mechanic1, mechanic2}. 

It is worth noting that the SSH model is noninteracting. By contrast, it is well known that real-life quantum systems are highly interacting in nature. Therefore, while the SSH model is helpful to highlight the role of topology in physics, it does not represent the full description of a typical quantum system. Unfortunately, incorporating interaction effects into the SSH model significantly increases its complexity. Indeed, solving quantum many-body problems is known to be a nontrivial problem due to their Hilbert space scaling exponentially (rather than linearly) with the system size. While some interacting variants of the SSH model have been studied in existing literature \cite{Liberto16, Salvo24,Koor22,Feng22,Yu20,Melo23}, sophisticated analytical and/or numerical techniques are typically employed, which can obscure their physical interpretations. 

In some cases, the effect of interaction can be approximated, at the mean field level, by nonlinearity. This amounts to turning the corresponding many-body interacting Hamiltonian of the target system by an effective single-body but nonlinear Hamiltonian. Here, a nonlinear Hamiltonian means that it is state dependent, thereby turning the otherwise linear Schrodinger equation into its nonlinear counterpart, e.g., the Gross-Pitaevskii equation \cite{Gross, Pitae}. Despite being a single-particle model, solving a nonlinear Schrodinger equation presents its own challenges. These include the breakdown of superposition principle and the difficulty in the ``diagonalization" procedure due to the state-dependence of the Hamiltonian. Nevertheless, solutions of an effective nonlinear system are typically easier to understand than their interacting counterparts. 

In view of the above, considering the effect of nonlinearity on an SSH model may, at least partially, shed light on the more elusive interacting SSH model. It is also worth noting that some type of nonlinearity may also arise naturally in some experimental realizations of the SSH model involving classical platforms, e.g., photonic waveguides \cite{photonic waveguides with Kerr nonlinearity}. In this case, studies of a nonlinear SSH model may also find some relevance to these experiments. Indeed, a type of nonlinear SSH model has in fact been introduced and examined in Ref.~\cite{2020-paper}. Its main results include the nonlinearity-induced modification of the Zak phase, the emergence of incomplete energy bands that form a loop structure, as well as the interplay between topology and soliton formation. 

In this work, we extend the nonlinear SSH model studied in Ref.~\cite{2020-paper}. Specifically, while Ref.~\cite{2020-paper} considers a uniform onsite nonlinearity, the present work investigates the effect of staggered onsite nonlinearity, whose amplitude depends on the sublattice of the SSH model. This results in more control over the nonlinear parameter, allowing for exploring the impact of nonlinearity on each sublattice independently. Our key findings, which will be elaborated below in detail, include a general expression of the nonlinearity-induced modification of the Zak phase (extending the applicability of that presented in Ref.~\cite{2020-paper} to a more general type of nonlinearity), nonlinearity-induced energy gap closing accompanied by discontinuity in the Zak phase, uncovering the dependence of each topological edge state on nonlinearity, the persistence of delocalized solutions at large nonlinear amplitudes in the case where they alternate between a positive value and a negative value, as well as the presence of a touching point between two energy solutions at very large nonlinear amplitudes that only shift (not disappear) in the presence of perturbation.     

This work is organized as follows. In Sec.~\ref{model}, we introduce the nonlinear Hamiltonian, and pose the (state-dependent) nonlinear eigenvalue problem, limiting ourselves to stationary solutions. In Sec.~\ref{Bloch}, we consider a special class of stationary solutions of the Bloch form, and use self-consistent analysis to solve the problem under Periodic Boundary Conditions (PBC), obtain energy bands, and study their geometric phases, then apply a dynamical stability analysis. Moreover, in section ~\ref{Real}, we present the numerical method used and our implementation to solve the problem in real space under Open Boundary Conditions (OBC), then show the main results obtained. Finally, in section ~\ref{conc} we summarize the paper with its key findings and discuss possible future directions.

\section{Model description}
\label{model}

We consider a nonlinear Rice-Mele lattice model \cite{Rice-Mele} that is described by a set of nonlinear Schrodinger equations of the form:
\begin{equation} \label{eq:real3.5}
i\frac{\dd{\psi_{S,m}}}{\dd{t}} = \sum_{S'=A,B} \sum_{m'=1}^N [\mathcal{H}(\psi)]_{S,S',m,m'} \psi_{S',m'} ,  
\end{equation}
where $S=A,B$, $\mathcal{H}$ is an effective nonlinear (state-dependent) Hamiltonian of the form
\begin{eqnarray}
\label{modelHamiltonian}
    \mathcal{H}(\psi) &=& \sum_{m=1}^{N-1} \left( \frac{J_2}{2} \sigma_+ \otimes |m\rangle \langle m+1| +h.c. \right) \nonumber \\
    && +\sum_{m=1}^N \left[J_1 \sigma_x + \left( \begin{array}{cc}
        g_A |\psi_{A,m}|^2 & 0 \\
        0 & g_B |\psi_{B,m}|^2
    \end{array} \right) \right.   \nonumber \\
    && \left. +v \sigma_z \right]\otimes |m\rangle \langle m| ,
\end{eqnarray}
$\sigma_s$ with $s=x,y,z$ are Pauli matrices in the sublattice degree of freedom, $\sigma_\pm = \sigma_x \pm \mathrm{i} \sigma_y$, $|m\rangle$ is a column vector of element $1$ on the $m$th row and zero elsewhere, $ \psi_{S,m} $ represents the wavefunction's amplitude at site with unit cell number $m$, sublattice $S\in \{A,B\}$, with $m \in \{1, 2, 3, ..., N \}$, and $N$ is a number that is large enough for the thermodynamic limit approximation $N \rightarrow \infty$ to apply. As to the model's parameters, $ J_{1}$ is the energy required to hop within the unit cell (the intracell hopping amplitude), $J_{2}$ is the energy required to hop between adjacent unit cells (the intercell hopping amplitude), $v$ is the onsite potential strength, $g_{A}~(g_{B})$ denotes the nonlinearity strength on sublattice $A~(B)$.

Throughout this paper, our focus is to seek stationary solutions of the above nonlinear time-independent Schrodinger equations, which satisfy
\begin{equation} \label{eq:real5}
\sum_{S'=A,B} \sum_{m'=1}^N [\mathcal{H}(\psi)]_{S,S',m,m'} \psi_{S',m'} = E \psi_{S,m} .
\end{equation}
While Eq. (\ref{eq:real5}) may look like a typical eigenvalue problem at first glance, it cannot be easily solved via the usual diagonalization technique due to the state-dependence of $\mathcal{H}(\psi)$. 

In the following, we shall solve Eq. (\ref{eq:real5}) by following two approaches. The first approach, which is discussed in detail in Sec.~\ref{Bloch}, focuses on a specific class of solutions (Bloch solutions) that obeys periodic boundary conditions (PBC). It simplifies Eq. (\ref{eq:real5}) by significantly reducing the number of degrees of freedom down from $2N$ to $2$, thereby allowing some semianalytical calculation to be carried out. The second approach, which is the main subject of Sec.~\ref{Real}, aims to access more general solutions of Eq. (\ref{eq:real5}) by numerically solving it via iterative self-consistent procedure. There, open boundary conditions (OBC) are further assumed, i.e.,
\begin{equation}
\label{eq:real6}
\psi_{B,~0} = \psi_{A,~N+1}  = 0 ,
\end{equation} 
so as to uncover the interplay between the formation of solitons and topological edge states.

\section{Bloch solution studies}
\label{Bloch}

For simplicity, we start by considering solutions that are of the Bloch form. Namely, such solutions can be written in the form
\begin{equation} \label{eq:rec1}
\begin{split}
\psi_{A,~j}~=~\varphi_{1}~e^{ikj}\\
\psi_{B,~j}~=~\varphi_{2}~e^{ikj}
\end{split}
\end{equation}
If Eq.~(\ref{eq:rec1}) is to be valid on a finite-sized lattice, periodic boundary conditions (PBC) are implied, i.e., $\psi_{A,N+1}=\psi_{A,1}$ and $\psi_{B,~0}=\psi_{B,~N}$, which restrict $k$ to take discrete values $k_j=\frac{2j\pi}{N}$ for $j=0,1,\cdots,N-1$. Note that $k$ becomes continuous in the thermodynamic limit $N\rightarrow \infty$, in which case the actual boundary conditions become irrelevant. In any case, by inserting Eq. (\ref{eq:rec1}) into  Eq. (\ref{eq:real5}), the stationary state equation becomes 
\begin{equation}\label{eq:rec2}
H(\Sigma,k) \Phi = E(k) \Phi
\end{equation}
where $H(\Sigma,k)$ is an effective $2\times 2$ state-dependent Hamiltonian of the form 
\begin{equation}\label{eq:rec3}
H(\Sigma,k) = h_{x}(k) \sigma_{x} + h_{y}(k) \sigma_{y} + h_{z}(\Sigma) \sigma_{z} + \frac{S(\Sigma)}{2} I ,
\end{equation}
where $\Phi$ is a pseudo-spinor $\Phi = 
\begin{pmatrix}
\varphi_{1}\\
\varphi_{2}
\end{pmatrix}$ , $\Sigma \equiv |\varphi_{2}|^2 - |\varphi_{1}|^2$, $\sigma_j$ are the $2\times 2$ Pauli matrices, $I$ is the $2\times 2$ identity matrix acting on $\Phi$, 
%\CH{(Please be consistent with the notation; either use $\varphi_j$ or $\Phi_j$ like in Eq. (5) only throughout the manuscript)}
\begin{eqnarray}
h_y(k) = J_{2} \sin{k} &,& h_x(k) = J_{1} + J_{2} \cos{k} ,   \nonumber \\ 
 h_{z}(\Sigma) &=& v + \frac{1}{4} \left[g_{A} - g_{B} - \Sigma (g_{A}+g_{B})\right] , \text{ and } \nonumber \\
 S(\Sigma) &=& \frac{1}{2}\left[(g_{B} - g_{A})\Sigma + g_{A} + g_{B}\right] . \label{eq:pardef}
\end{eqnarray}
If $g_A=g_B=0$, Eq.~(\ref{eq:rec3}) reduces to the Rice-Mele Hamiltonian in the momentum space \cite{Rice-Mele}. It further reduces to the momentum space Hamiltonian of the SSH model \cite{SSH} if $v=0$, which is known to be topologically trivial (nontrivial) for $J_1>J_2$ ($J_2>J_1$). The eigenvalue equation for both models is easily solved as $E_\pm =  \pm \sqrt{h_{x}^{2}+h_{y}^{2}+h_{z}^{2}}$ due to the well-known algebra of the Pauli matrices. 

At nonzero nonlinear strengths $g_A$ and $g_B$, Eq.~(\ref{eq:rec3}) can similarly be solved by first writing the energy solutions as
\begin{equation} \label{eq:rec4}
E_\pm = \frac{S(\Sigma)}{2} \pm \sqrt{h_{x}^{2}+h_{y}^{2}+h_{z}^{2}(\Sigma)} .
\end{equation}
However, $E_\pm$ are not fully determined from Eq.~(\ref{eq:rec4}) as the right hand side still depends on the unknown pseudo-spinor elements $\varphi_1$ and $\varphi_2$ through $\Sigma$. To obtain the closed expression for $E_{1,2}$, we may use the fact that a pseudo-spinor state can be written in the Bloch sphere representation as
\begin{equation}
 \Phi = 
\begin{pmatrix}
\varphi_{1}\\
\varphi_{2}
\end{pmatrix}  = 
\begin{pmatrix}
\cos\frac{\theta}{2}\\
e^{i\phi} \sin\frac{\theta}{2}
\end{pmatrix} ,
\end{equation}
where $\theta \in [0, ~\pi]$, and $\phi \in [0,~2\pi)$. In particular, 
\begin{equation}
 \Sigma = -\cos\theta = -\frac{h_z(\Sigma)}{\sqrt{h_x^2+h_y^2+h_z^2(\Sigma)}} ,   
\end{equation}
which, using the explicit form of $h_z(\Sigma)$, can be turned into a quartic polynomial in $\Sigma$. The physical roots of such a polynomial, i.e., which are real-valued and satisfy $|\Sigma| \leq1$, can then be plugged in back to Eq.~(\ref{eq:rec4}) to obtain the complete energy spectrum.

\subsection{Energy spectrum}
\label{Energy}

\begin{figure*}[!]               % use [t] or [b] for top/bottom placement
    \centering
    %--- Row 1 -------------------------------------------------
    \begin{subfigure}{0.48\linewidth}
        \includegraphics[width=\linewidth]{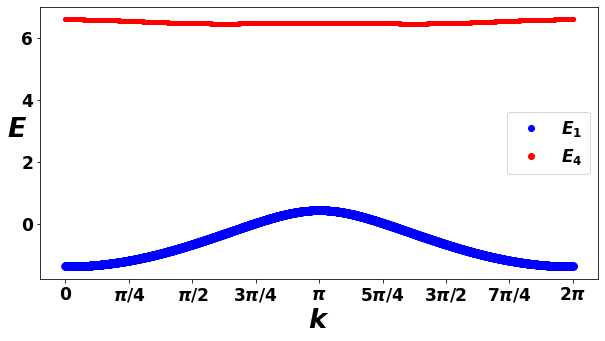}
        \caption{}
        \label{fig:A}
    \end{subfigure}\hfill
    \begin{subfigure}{0.48\linewidth}
        \includegraphics[width=\linewidth]{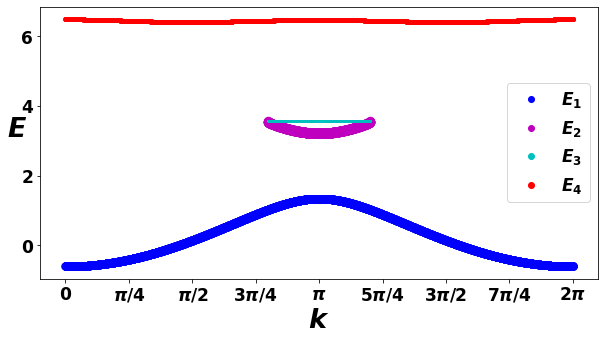}
        \caption{}
        \label{fig:B}
    \end{subfigure}
    
    \vspace{0.8em}
    
    \begin{subfigure}{0.48\linewidth}
        \includegraphics[width=\linewidth]{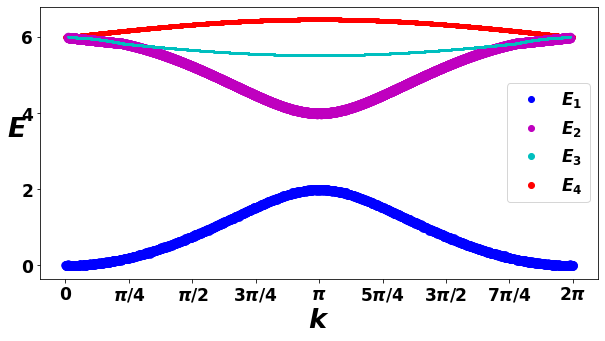}
        \caption{}
        \label{fig:C}
    \end{subfigure}\hfill
     % small vertical gap between rows
    %--- Row 2 -------------------------------------------------
    \begin{subfigure}{0.48\linewidth}
        \includegraphics[width=\linewidth]{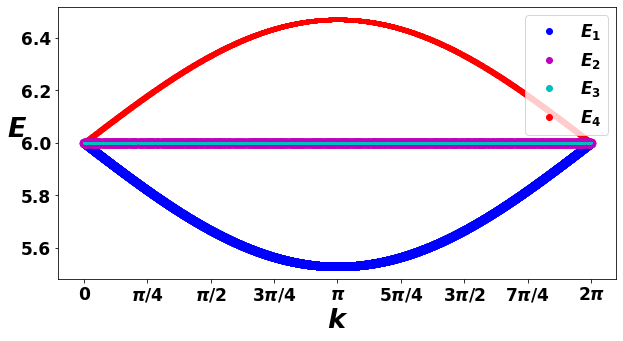}
        \caption{}
        \label{fig:D}
    \end{subfigure}
    
    \vspace{0.8em}
    
    \begin{subfigure}{0.48\linewidth}
        \includegraphics[width=\linewidth]{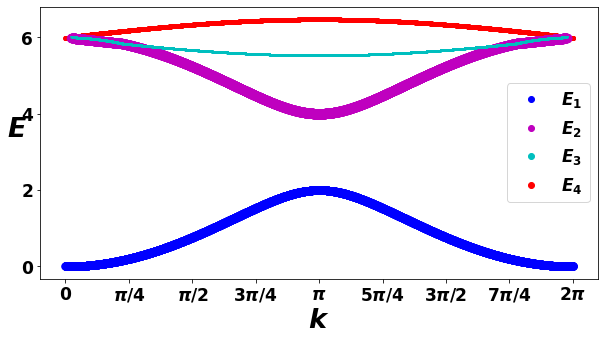}
        \caption{}
        \label{fig:E}
    \end{subfigure}\hfill
    \begin{subfigure}{0.48\linewidth}
        \includegraphics[width=\linewidth]{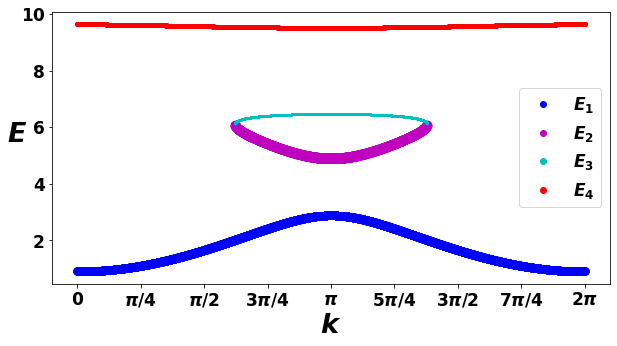}
        \caption{}
        \label{fig:F}
    \end{subfigure}

    %--- Overall caption -----------------------------------------
    \caption{Plots of energy bands (E) vs quasimomentum (k) for $J_{1}=1~,~J_{2}=2~, ~v=0.5~,~g_{B}=7~$, as nonlinearity strength $g_{A}$ increases, for:  (a) $g_{A}=1$ , (b) $g_{A}=3$ , (c) $g_{A}=4.99999$ , (d) $g_{A}=5$ (topological phase transition point) , (e) $g_{A}=5.0001$ , and (f) $g_{A}=9$.}
    \label{Fig.3.5}
\end{figure*}

Following the argument presented above and by leaving the mathematical detail in Appendix.~(\ref{Quartic}), we obtain the quartic equation
\begin{equation} \label{quartic_poly}
\small
\begin{split}
&x^{4} \left[(g_{A})^{2} + (g_{B})^{2} + 2g_{A}~g_{B} \right] + x^{3} [4~v~(g_{A}+g_{B}) - 4~g_{A}~g_{B}\\
&- (g_{A})^{2} -3~(g_{B})^{2}] + x^{2} [4~(J_{1}^{2} ~+ J_{2}^{2}~+v^{2}~+ 2~J_{1}~J_{2}~\cos{k})\\
&+3~g_{B}^{2} ~+ 2g_{A}~g_{B} - 4v~g_{A} -8v~g_{B}] + x~[ 4v~g_{B} -4~(J_{1}^{2} ~+ J_{2}^{2}~\\
&+v^{2}~+ 2~J_{1}~J_{2}~\cos{k}) -g_{B}^{2}] + J_{1}^{2} ~+ J_{2}^{2}~+ 2~J_{1}~J_{2}~\cos{k} = 0
\end{split}
\end{equation} where $x\equiv\cos^{2}\frac{\theta}{2}$ for notational convenience. Throughout this manuscript, thermodynamic limit is implied, so that $k \in [0,~2\pi)$. The above quartic equation can then be solved numerically for a given $k$ value, and only real roots $x \in \mathbb{R}$ with  $0 \leq x \leq 1$ are selected in the computation of the corresponding energy solution, i.e., via Eq.~(\ref{eq:rec4}). As such a quartic equation supports four roots, at most four energy solutions are obtained for each value of $k$. However, since some of these roots are not physical, i.e., either they are complex-valued or outside the interval $[0,1]$, some values of $k$ exhibit less than four energy solutions. The latter leads to the presence of some ``incomplete energy bands" within the Brillouin zone $k\in [0,2\pi)$, which have no linear counterparts. Such incomplete energy bands typically form a loop structure \cite{loop1, loop2, loop3, loop4, loop5}, as numerically demonstrated in Fig.~\ref{Fig.3.5}(b).

Figure~\ref{Fig.3.5}(a)-(d) further shows that, as one nonlinearity strength ($g_A$ or $g_B$) increases, the region which supports four physical roots (corresponding to four energy solutions) enlarges within the Brillouin zone. In particular, at a sufficiently large nonlinearity strength, the loop structure extends throughout the whole Brillouin zone, and the system exhibits four energy bands in total (Fig.~\ref{Fig.3.5}(c)). Interestingly, at a specific value of the nonlinearity strengths, e.g., at $g_A=5$ and $g_B=7$ in our numerics, the energy bands exhibit abrupt structural changes and become gapless (see and compare Fig.~\ref{Fig.3.5}(c) and Fig.~\ref{Fig.3.5}(d)). 

\begin{figure}[!]
    %\centering
    \includegraphics[width=1\linewidth]{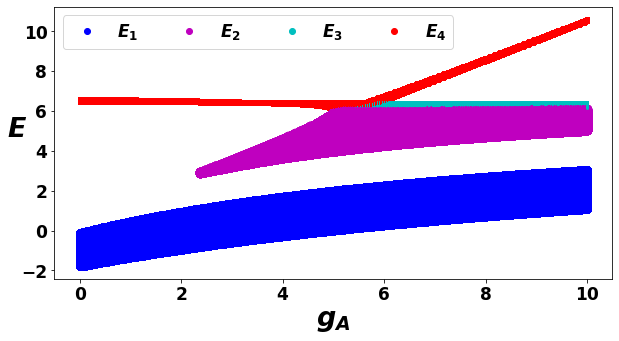}
    \caption{Energy vs parameter $g_{A}$ for $J_{1}=1~,~J_{2}=2~, ~v=0.5~,~g_{B}=7~$ in the Bloch solution studies. Each value of $g_A$ contains the energy solutions from all possible $k$ values.}
    \label{Fig3}
\end{figure}

To get more insight into the structure of the energy bands, we plot in Fig.~\ref{Fig3} the combined energy eigenvalues for all values of $k$ in the Brillouin zone as a function of the nonlinearity strength $g_A$. At small values of $g_A$, there are initially only two energy bands in the system. As $g_A$ reaches some critical value (around $g_A\approx 2.4$ for a given set of parameter values used in our numerics), a pair of additional bands start to emerge, the width of which increases with $g_A$. The bandwidth stops increasing at another critical value of $g_A$ ($g_A=5$ for a given set of parameter values used in our numerics), which is marked by the gap closing between the upper three bands in Fig.~\ref{Fig3}. Further increase in $g_A$ beyond this critical value leads to a different behavior in the upper band whereby its corresponding energy also increases almost linearly in $g_A$ rather than being at almost a constant value. This suggests that such a critical value corresponds to some kind of phase transition. In the next section, we investigate a possible topological origin of the phase transition by defining and evaluating the system's ``nonlinear Zak phase".

\begin{figure}[!]
    %\centering
    \includegraphics[width=1\linewidth]{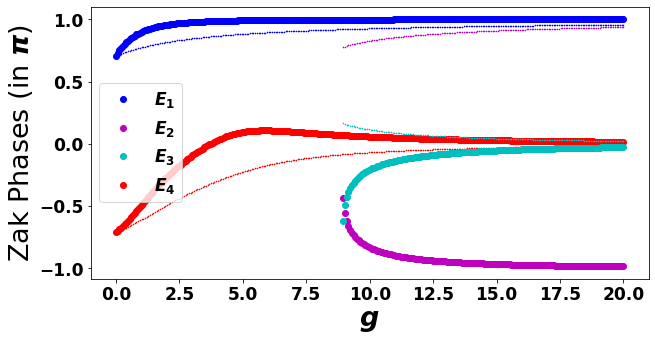}
    \caption{Reproduction of Zak phases plot obtained in Fig.~3 of Ref.~\cite{2020-paper} using the general nonlinear Zak phase expression in Eq.~(\ref{eq:a15}). The general nonlinear Zak phase was plotted in thick marks, whereas the conventional Zak phase was plotted in thin marks. The system parameters are chosen as $g_A=g_B=g$ for $J_{1}=1~,~J_{2}=2~, ~v=0.5~$.}  
    \label{Zak-Phases-Reproduction}
\end{figure}
%\begin{figure}[!]
%    \centering
%    \includegraphics[width=1\linewidth]{(J1=1, J2= 2, v=0.5, gB=9, gA=-20 to 0).png}
%    \caption{Energy vs parameter $g_{A}$ for $J_{1}=1~,~J_{2}=2~, ~v=0.5~,~g_{B}=9~$}
%    \label{Fig5}
%\end{figure}

%\begin{figure}[!]
%    \centering
%    \includegraphics[width=1\linewidth]{(J1=1, J2= 2, v=0.5, gA=-10, gB=9).png}
%    \caption{Energy(E) vs quasimomentum (k) for $J_{1}=1~,~J_{2}=2~, ~v=0.5~,~g_{A}=-10,~~g_{B}=9~$}
%    \label{Fig5.5}
%\end{figure}

\subsection{Nonlinear Zak phase}
\label{Zak}

It is well-known that the topology of the linear SSH model is distinguished by the corresponding Zak phase Ref.~\cite{ZakPhase}. In particular, a Zak phase of zero ($\pi$) corresponds to the trivial (nontrivial) phase that is marked by the absence (presence) of zero energy edge modes. In the presence of onsite nonlinearity, Ref.~\cite{2020-paper} found that the definition of Zak phase requires some modification due to the geometric-dependent contribution from the dynamical phase. That is, the conventional Zak phase formula, i.e., 
\begin{equation}
    \gamma_{\rm lin}=\int_0^{2\pi} i \Phi^{\dagger}\frac{\dd{\Phi}}{\dd{k}} dk  
\end{equation}
generally does not provide the complete geometric description of the adiabatic process in the nonlinear system. For the nonlinear system considered in Ref.~\cite{2020-paper}, which corresponds to setting $g_{A}=g_{B}$ in Eq.~(\ref{eq:pardef}), the modified Zak phase (termed ``nonlinear Zak phase" in Ref.~\cite{2020-paper}) can be significantly different from the conventional Zak phase. Remarkably, at sufficiently high nonlinearity, nonlinear Zak phases of all nonlinear energy bands were found to sum up to a quantized quantity (0 modulo 2$\pi$), whilst the sum of the corresponding conventional Zak phases is not quantized \cite{2020-paper}. In addition, the nonlinear Zak phase for each nonlinear energy band, although not quantized, is closer than conventional Zak phase to either $0$ or $\pi$. In Ref.~\cite{2020-paper}), these features are presented as evidence that the nonlinear Zak phase is indeed the more appropriate quantity to consider than the conventional Zak phase in the nonlinear setting.

In view of the above, we generalize the derivation of the nonlinear Zak phase presented in Ref.~\cite{2020-paper} so as to obtain a more general expression that works for any real-valued well-behaved functions $h_{x}(k) , h_{y}(k)$, and for any real $C^{1}$ functions $h_{z}(\Sigma)$ and $ S(\Sigma)$, the derivation of which is detailed in Appendix~\ref{Appen-Zak}. For the specific nonlinear model considered in this work, this nonlinear Zak phase takes the explicit form,

\begin{equation} \label{Nonlinear-Zak-Phase}
    \gamma_{\rm Nonlin}=i\int_0^{2\pi} \epsilon \alpha_{1}~ dk  ,
\end{equation}
where
\begin{widetext}
\begin{equation} \label{eq:rec7}
\epsilon \alpha_{1} = 
\frac{i \Phi^{\dagger}\frac{\dd{\Phi}}{\dd{t}} + 4 i \cos\frac{\theta}{2} \frac{\dd{\varphi_{1}}}{\dd{t}} \left( \frac{\frac{1}{4} \left[g_{B} - g_{A} + \Sigma (g_{A} + g_{B}) \right]}{ -E + v + \frac{g_{A}}{2} (1-\Sigma)  - \sqrt{h_{x}^{2} + h_{y}^{2}} \cot\frac{\theta}{2} + 2 g_{A} \cos^{2}\frac{\theta}{2}} \right)}{1 + 4 \cos^{2}\left(\frac{\theta}{2}\right) \left( \frac{\frac{1}{4}\left[g_{B} - g_{A} + \Sigma (g_{A} + g_{B}) \right]}{-E + v + \frac{g_{A}}{2} (1-\Sigma)  - \sqrt{h_{x}^{2} + h_{y}^{2}} \cot\frac{\theta}{2} + 2 g_{A} \cos^{2}\frac{\theta}{2}} \right)} .
\end{equation}     
\end{widetext}

\begin{figure}[!]
    \centering
    \includegraphics[width=1\linewidth]{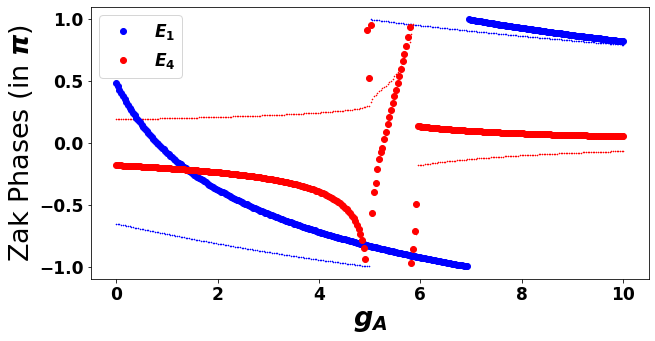}
    \caption{General nonlinear Zak phase (in bold) and Conventional Zak phase (in thin) vs parameter $g_{A}$ for $J_{1}=1~,~J_{2}=2~, ~v=0.5~,~g_{B}=7~$.}
    \label{Fig4}
\end{figure}

In the linear limit, i.e., $g_{A} = g_{B}=0$, the fraction within $(\cdots)$ vanishes and the expression reduces to the conventional Zak phase $\gamma_{\rm lin}$ as expected. Meanwhile, in the limit $g_{A} = g_{B}=g$, we managed to successfully reproduce the Zak phases plot (up to an unimportant minus sign factor) presented in Fig.~3 of \cite{2020-paper} (see Fig.~\ref{Zak-Phases-Reproduction} below). Together, these checks verify the validity of Eq.~(\ref{eq:rec7}).  
%\CH{(Please move Fig.~\ref{Zak-Phases-Reproduction} earlier, right after Fig. 2)}
In Fig.~\ref{Fig4}, we plot both the numerically calculated nonlinear Zak phase and the corresponding conventional Zak phase for the outer two bands at varying $g_A$ (the other system parameters are fixed to the values used in Fig.~\ref{Fig3}). Interestingly, a jump-discontinuity is observed in both Zak phases at $g_{A}=5$, i.e., at precisely the phase transition point identified in Fig.~\ref{Fig3}. This confirms the topological origin of such a phase transition, i.e., in the same spirit as the topological phase transition in the linear SSH model that is marked by a jump-discontinuity in its Zak phases. 

%When we studied the geometric phases (general nonlinear Zak phase in Eq. (\ref{eq:rec7}) and conventional Zak phase) of the outermost nonlinear energy bands $(E_1 ~, E_{4})$ plotted in Fig. \ref{Fig3}, a jump-discontinuity was observed at $g_{A}=5$ (see Fig. \ref{Fig4}), which happened to coincide with energy gap-closing as shown in Fig. \ref{Fig3}. Such observations strongly suggest that at $g_{A}=5$ one have a nonlinearity-induced topological phase transition. However, when Zak phases were computed for the energy bands in Fig. \ref{Fig5}) (corresponding to $g_{A}$ and $g_{B}$ having different signs), we see from the plot in Fig. \ref{Fig6} that although the conventional Zak phase is not generally quantized, at sufficiently high nonlinearity strength, the general nonlinear Zak phase approaches quantized value. 

%\begin{figure}[!]
%    \centering
%    \includegraphics[width=1\linewidth]{(1, 2, 0.5, gB=9, gA neg).png}
%    \caption{General nonlinear Zak phase (in bold), and Conventional Zak phase (in thin) vs parameter $g_{A}$ for $J_{1}=1~,~J_{2}=2~, ~v=0.5~,~g_{B}=9~$}
%    \label{Fig6}
%\end{figure}

\subsection{Dynamical stability of energy bands}
\label{Stability}

In nonlinear systems, due to the lack of superposition principle, a state prepared close to but not exactly in a stationary state may either remain close to the stationary state or become infinitely far apart from the stationary state in the long run. When any arbitrary state is in the vicinity of a stationary state that satisfies the former, the stationary state is said to be dynamically stable. Otherwise, it is dynamically unstable. The dynamical stability of a stationary state can be determined by inspecting the spectral structure of the $\mathcal{L}$ matrix, the exact definition of which is presented in Eq.~(\ref{eq:ab6}) of Appendix~\ref{ch2,3}. 

Intuitively, the $\mathcal{L}$ matrix describes the first-order difference in the nonlinear Hamiltonian under an exact stationary state and that under another state close to it. The Schrodinger equation associated with the $\mathcal{L}$ matrix then effectively quantifies the time evolution of the state difference. As the $\mathcal{L}$ matrix is linear, its complete eigenvalues and the corresponding eigenvectors can be easily obtained. However, as $\mathcal{L}$ is generally non-Hermitian, its eigenvalues can be complex. The presence of complex eigenvalues is usually associated with dynamical instability, as they generate factors that grow exponentially with time. Conversely, the absence of complex eigenvalues is associated with dynamical stability. 

As detailed in Appendix~(\ref{ch2,3}), we obtain
\begin{equation}\label{ch4,4.2,1}
\mathcal{L} = 
\scalebox{0.5}{%
$
\begin{pmatrix}
v+2g_{A}|\psi_{1}^{(0)}|^2-E & h_{x}(k)-ih_{y}(k) & g_{A}\left(\psi_{1}^{(0)}\right)^2 & 0\\
h_{x}(k)+ih_{y}(k) & -v+2g_{B}|\psi_{2}^{(0)}|^2-E & 0 & g_{B}\left(\psi_{2}^{(0)}\right)^2\\
-g_{A}\left(\psi_{1}^{(0)~\ast}\right)^2 & 0 & -\left(v+2g_{A}|\psi_{1}^{(0)}|^2-E\right) & -\left[h_{x}(k)+ih_{y}(k)\right]\\
0 & -g_{B}\left(\psi_{2}^{(0)~\ast}\right)^2 & -\left[h_{x}(k)-ih_{y}(k)\right] & -\left(-v+2g_{B}|\psi_{2}^{(0)}|^2-E\right)
\end{pmatrix}$}
\end{equation}
associated with any stationary state $\Psi=\Psi^{(0)}$. We then define the quantity
\begin{equation}\label{ch4,4.2,2}
\max{[|\Im{(\lambda_{n})}|]} ~, ~ \forall n ,
\end{equation} 
where $\Im(\cdots)$ stands for the imaginary part of $(\cdots)$ and $\lambda_n$ is the $n$th eigenvalue of $\mathcal{L}$. The state is dynamically unstable (stable) if $\max{[|\Im{(\lambda_{n})}|]}\neq 0$ ($\max{[|\Im{(\lambda_{n})}|]}=0$). 

\begin{figure*}[!]               % use [t] or [b] for top/bottom placement
    \centering
    %--- Row 1 -------------------------------------------------
    \begin{subfigure}{0.48\linewidth}
        \includegraphics[width=\linewidth]{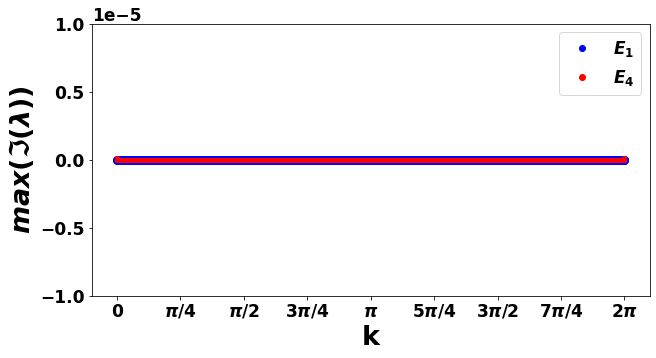}
        \caption{}
        \label{plotch4,5}
    \end{subfigure}\hfill
    \begin{subfigure}{0.48\linewidth}
        \includegraphics[width=\linewidth]{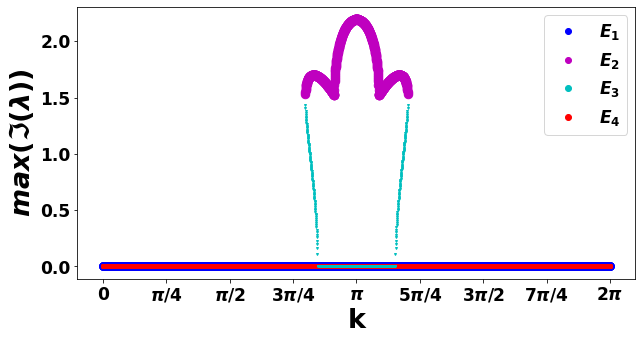}
        \caption{}
        \label{plotch4,7}
    \end{subfigure}
    
    \vspace{0.8em}
    
    \begin{subfigure}{0.48\linewidth}
        \includegraphics[width=\linewidth]{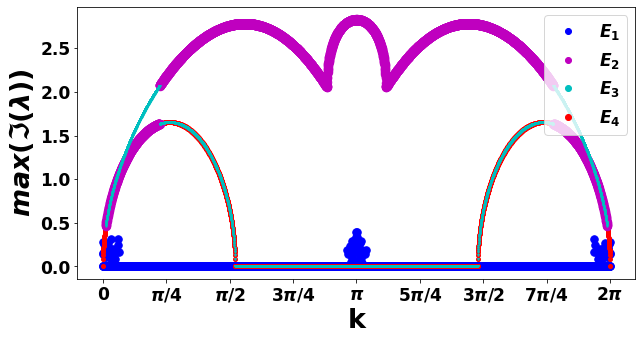}
        \caption{}
        \label{Stab:C}
    \end{subfigure}\hfill
     % small vertical gap between rows
    %--- Row 2 -------------------------------------------------
    \begin{subfigure}{0.48\linewidth}
        \includegraphics[width=\linewidth]{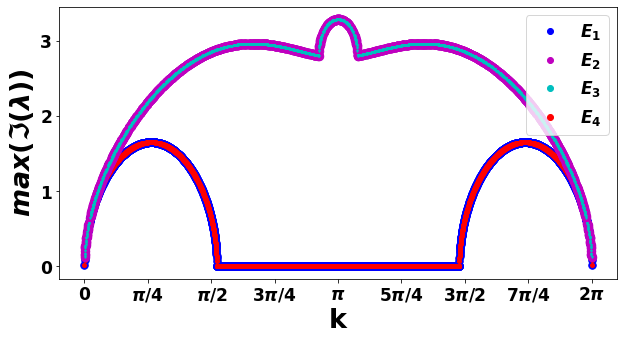}
        \caption{}
        \label{fig:D}
    \end{subfigure}
    
    \vspace{0.8em}
    
    \begin{subfigure}{0.48\linewidth}
        \includegraphics[width=\linewidth]{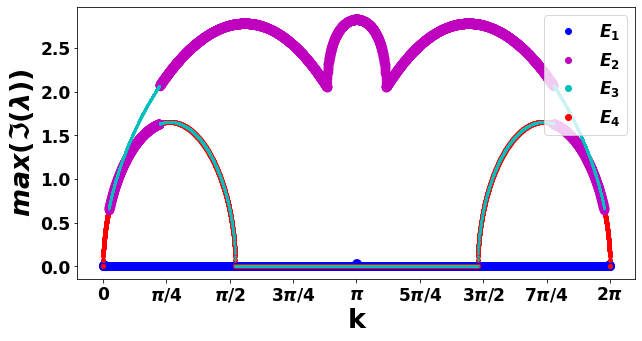}
        \caption{}
        \label{fig:E}
    \end{subfigure}\hfill
    \begin{subfigure}{0.48\linewidth}
        \includegraphics[width=\linewidth]{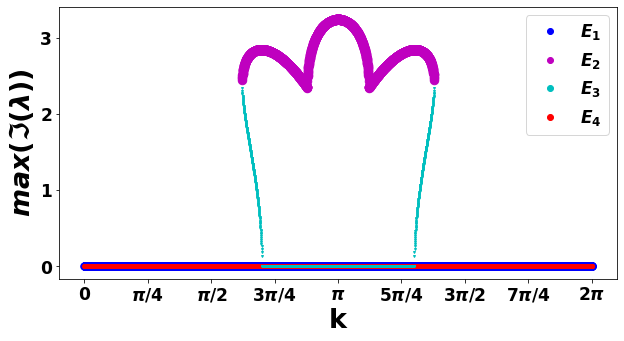}
        \caption{}
        \label{fig:F}
    \end{subfigure}

    %--- Overall caption -----------------------------------------
    \caption{Dynamical stability analysis of energy bands corresponding to the parameters used in Fig.~\ref{Fig.3.5}, i.e., for $J_{1}=1~,~J_{2}=2~, ~v=0.5~,~~g_{B}=7~$, as nonlinearity strength $g_{A}$ increases, for:  (a) $g_{A}=1$ , (b) $g_{A}=3$ , (c) $g_{A}=4.99999$ , (d) $g_{A}=5$ (Topological Phase Transition Point) , (e) $g_{A}=5.0001$ , and (f) $g_{A}=9$.}
    \label{Stab.3.5}
\end{figure*}

Our results are summarized in Fig.~\ref{Stab.3.5}, which reveals that the two outer bands are generically stable. Meanwhile, the two additional bands that form a loop structure are always found to be dynamically unstable. Near the phase transition, the outer band also develops some dynamical instability at $k$ values near $0$ and $2\pi$. This dynamical instability region could be attributed to the fact that the outer band touches the two additional bands at $k=0,2\pi$ at the phase transition point (see Fig.~\ref{Fig.3.5}(c,e)). 

%\begin{figure}[!]
%    \centering
%    \includegraphics[width=1\linewidth]{Stability (J1=1, J2= 2, v=0.5, gA=-10, gB=1).png}
%    \caption{Dynamical stability analysis of energy bands for $J_{1}=1~,~J_{2}=2~, ~v=0.5~,~g_{A}=-10,~~g_{B}=1~$}
%    \label{plotch4,17}
%\end{figure}

%\begin{figure}[!]
%    \centering
%    \includegraphics[width=1\linewidth]{Stability plot(J1=1, J2= 2, v=0.5, gA=-10, gB=9).png}
%    \caption{Dynamical stability analysis of energy bands for $J_{1}=1~,~J_{2}=2~, ~v=0.5~,~g_{A}=-10,~~g_{B}=9~$}
%    \label{plotch4,19}
%\end{figure}

\section{Real space studies}
\label{Real}

\subsection{Numerical method and implementation}
\label{Numerical}
%\CH{(If this is a known method, then a proper citation should be made)}

We now relax the Bloch solution assumption made in the previous section and attempt to obtain a more general set of solutions. This time, we solve Eq.~(\ref{eq:real3.5}) under OBC, so as to ensure that the obtained solutions are distinct from what we previously found under Bloch solution assumption. To this end, we numerically employ the Self-Consistent Field (SCF) iterative method \cite{SCF method}. The iteration for a state-dependent nonlinear Hamiltonian $H_{OBC}$ from state $\Psi_{n}$ to state $\Psi_{n+1}$, where $n$ labels the iteration step, goes as follows:

\begin{enumerate}
  \item Evaluate $H_n\equiv H_{OBC} (\Psi_{n})$
  \item Find the eigenstates $\Phi_{i}$ of $H_{n}~, ~i= 1, 2, ..., 2N $ .
  \item Select a new state $\Psi_{n+1} $ to be the eigenstate $\Phi_{j}$ that has the maximum inner product with the previous $\Psi_{n}$ state, i.e. $\Phi_{j}$ that satisfy $|\Psi_{n}^{\dagger}~\Phi_{j}| > 1-s $ ~where $s$ is the tolerance threshold, which is set to $ s=10^{-10}$ in our numerics. 
\end{enumerate}
%Note that the system Eq.~(\ref{eq:real5}) has 5 parameters $(J_{1}~, ~J_{2}~, ~v~,~g_{A}~,~g_{B})$. At the beginning of each iteration process, an energy eigenstate $\Psi_{0}$, from the corresponding linear model (i.e. the system with the following parameters $(J_{1}~, ~J_{2}~, ~v~,~0~,~0)$, which is exactly solvable), is selected as an initial trial state $\Psi_{0}$. 

In our study, all energy eigenstates from the corresponding linear model (corresponding to $g_A=g_B=0$) were selected as initial trial states in the above iterative procedure. However, only nonlinear stationary solutions that converged sufficiently quickly (up to the convergence threshold of 150 iterations) were collected and plotted later. It should therefore be emphasized that this implementation of the numerical method does not exhaustively capture all possible solutions of our nonlinear system, but only numerically stable ones. Nevertheless, as presented below, the obtained solutions already exhibit rich properties that capture the effect of staggered nonlinearity considered in our model.

%Yet, to ensure that we consider as many converging solutions as possible, we studied the convergence of representative solutions and found them to converge in less than 20 iterations, after that we selected the convergence threshold to be 150 iterations. 

We now define the Inverse Participation Ratio (IPR) \cite{IPR}: %\CH{(Cite at least one paper that uses this metric)}:
\begin{equation*} 
IPR\left( \Psi\right) = \frac{1}{\sum_{S=A,B}\sum_{m=1}^{N} |\psi_{S,m}|^4}
\end{equation*}
where $\psi_{S,m}$ is the wavefunction amplitude at site with unit cell number $m$, sublattice $S$. Note that the IPR for a localized solution is around 1, and for a delocalized solution $\rightarrow \infty$. In this paper, we use the term ``soliton" loosely to refer to any localized state. To further distinguish between the different types of solutions, the four following tests are made (in their exact order) in our numerics, i.e., 

\begin{enumerate}
  \item If \begin{equation*} 
0.15 < |\psi_{A,1}|^2 + |\psi_{B,1}|^2 + |\psi_{A,2}|^2 + |\psi_{B,2}|^2
\end{equation*} is satisfied, then the solution is localized near the left edge and is plotted in cyan.
  \item Otherwise, if \begin{equation*} 
0.15 < |\psi_{A,N-1}|^2 + |\psi_{B,N-2}|^2 + |\psi_{A,N}|^2 + |\psi_{B,N}|^2  
\end{equation*} is satisfied, then the solution is localized near the right edge and is plotted in red. 
  \item Otherwise, if \begin{equation*} 
0.85 ~< IPR\left( \Psi\right) < ~5
\end{equation*} is satisfied, then the solution is localized in the bulk and is plotted in black. 
  \item If none of the above conditions are satisfied, then the solution is a delocalized bulk state, i.e., not a soliton, and is plotted in gray.
\end{enumerate}
Note that the first two conditions are not mutually exclusive, so a state can, in theory, satisfy the first two conditions simultaneously. For such a case in our algorithm, the state will be labeled left edge state, since point 1 was tested first. While specifying a fifth color for states that simultaneously satisfy both conditions could be an option that avoids such an ambiguity, it was not implemented in our numerics due to the fact that, by considering a sufficiently long lattice and directly observing all the relevant wavefunction's profiles (not shown), all edge solitons we obtain are either localized to the left or right edge, but not both. 

\subsection{Real space results}
\label{real-results}

\begin{figure}[!]
    \centering
    \includegraphics[width=1\linewidth]{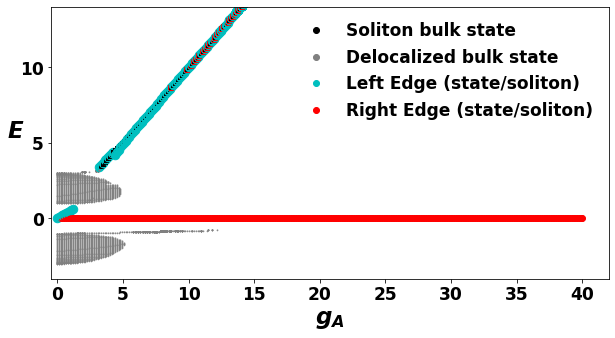}
    \caption{Energy vs parameter $g_{A}$ for $J_{1}=1~,~J_{2}=2~, ~v=0.0001~,~g_{B}=0~$ in the real space studies. The lattice size is taken as 100 sites here and in the subsequent figures involving our real space results.}
    \label{Fig1}
\end{figure}

We will now present the main results of our real space studies which involve different sign variations of $g_A$ and $g_B$. The lattice's size throughout the real-space study was set to be of $100$ sites in total, which is large enough for the thermodynamical limit to apply. In the case where $g_{A}$ and $g_{B}$ are both positive, it was found that in the topologically nontrivial regime ($J_1 < J_2$) and at sufficiently high nonlinearities, the energy of the left (right) edge state depends only on $g_{A}$ ($g_{B}$) and is constant with respect to $g_{B}$ ($g_{A}$) (see Figs. \ref{Fig1}, \ref{ch4,1}, \ref{ch4,1.5}). Moreover, the existence of only one edge state was found in the model for a particular set of values of a sublattice-dependent nonlinearity strength. For instance, note the disappearance of the left edge state in the region around $g_{A}=3$ in Fig. \ref{Fig1}, where the nonexistence of the left edge state was also verified after checking the wave profiles of all solutions near $g_{A}=3$. These observations could potentially have significant roles in quantum-state-transfer-like applications. Specifically, by starting with a set of system parameters that support only a single edge state, it can be utilized to simulate the encoding of quantum information. A series of adiabatic variations of system parameters could then, in principle, be devised to systematically move this edge state from one end to the other, effectively simulating a quantum state transfer process \cite{Quantum-State-Transfer1, Quantum-State-Transfer2}.

\begin{figure}
    \centering
    \includegraphics[width=1\linewidth]{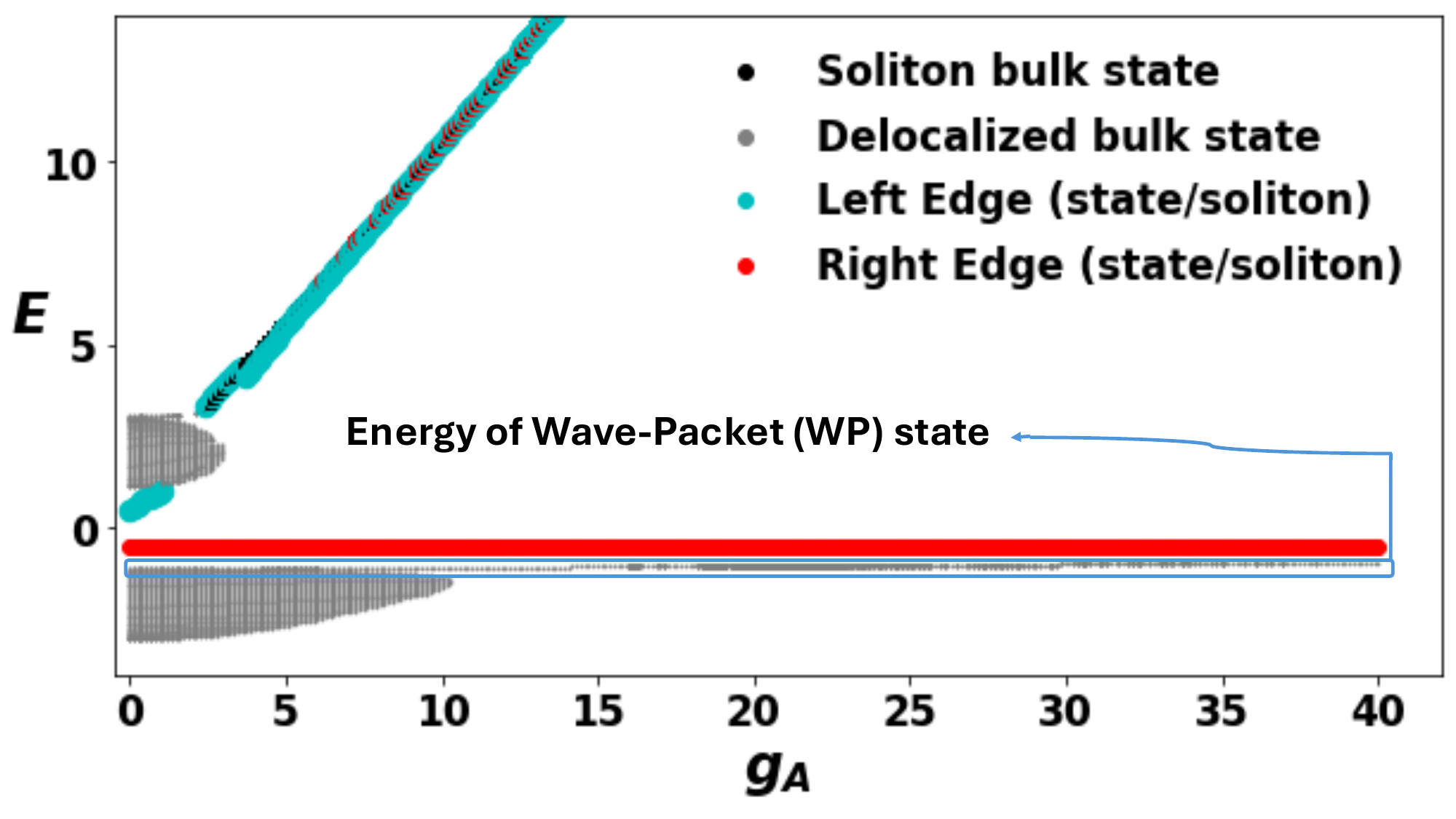}
    \caption{Energy vs parameter $g_{A}$ for $J_{1}=1~,~J_{2}=2~, ~v=0.5~,~g_{B}=0~$ in the real space studies.}
    \label{ch4,1}
\end{figure}

\begin{figure}
    \centering
    \includegraphics[width=1\linewidth]{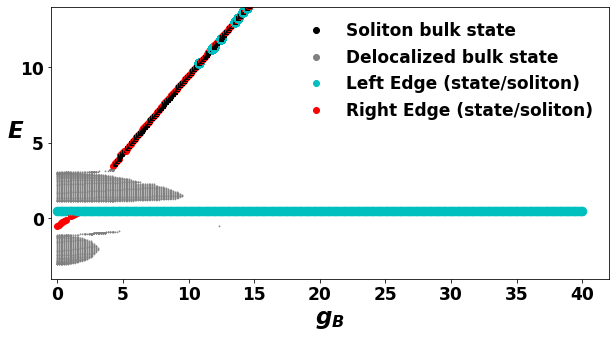}
    \caption{Energy vs parameter $g_{B}$ for $J_{1}=1~,~J_{2}=2~, ~v=0.5~,~g_{A}=0~$ in the real space studies.}
    \label{ch4,1.5}
\end{figure}

\begin{figure*}[!]
\begin{subfigure}{0.50\textwidth}
  \centering
  \includegraphics[width=1\linewidth]{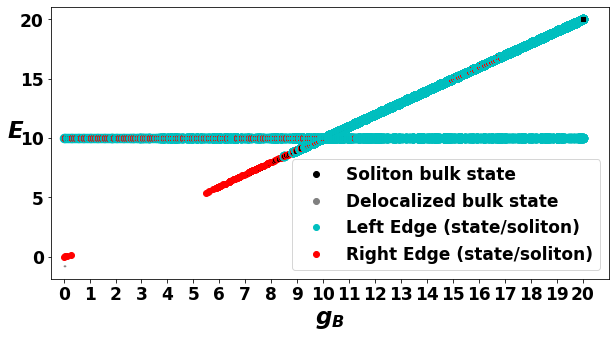}
  \caption{$~v=0.0001~$}
  \label{Fig2}
\end{subfigure}%
\begin{subfigure}{0.50\textwidth}
  \centering
  \includegraphics[width=1\linewidth]{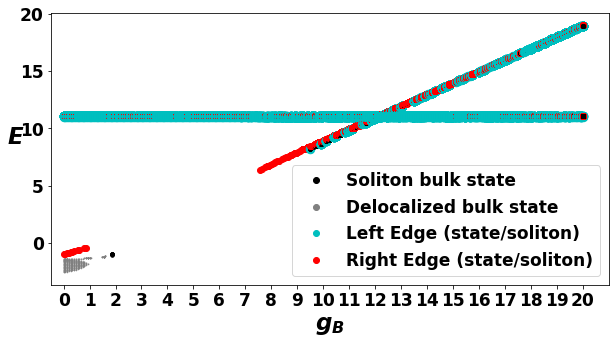}
  \caption{$~v=1~$}
  \label{Fig2.4}
\end{subfigure}
\caption{Plots of Energy vs parameter $g_{B}$ for $J_{1}=1~,~J_{2}=2~,~g_{A}=10~$ in the real space studies.}
%\label{ch4,2}
\end{figure*}

At sufficiently high nonlinearities, i.e., large values of $g_A$ and/or $g_B$, Fig. \ref{Fig1} also reveals that delocalized solutions cease to exist. In this case, by fixing one nonlinearity parameter while varying the other, we note the presence of a touching point between the two energy bands as shown in Fig. \ref{Fig2}. This touching point was found to only shift (and not disappear) under variations in the onsite potential strength $v$ (see Fig. \ref{Fig2.4}), which could indicate a topological origin analogous to gapless topological phases observed in Weyl semimetal for instance \cite{Weyl1, Weyl2, Weyl3, Weyl4, Weyl5, Weyl6, Weyl7, Weyl8}. Such a touching point was not observed in the case of uniform nonlinearity, i.e., $g_{A}=g_{B}$, even at large nonlinearity values \cite{2020-paper}.

\begin{figure}[!]
    \centering
    \includegraphics[width=1\linewidth]{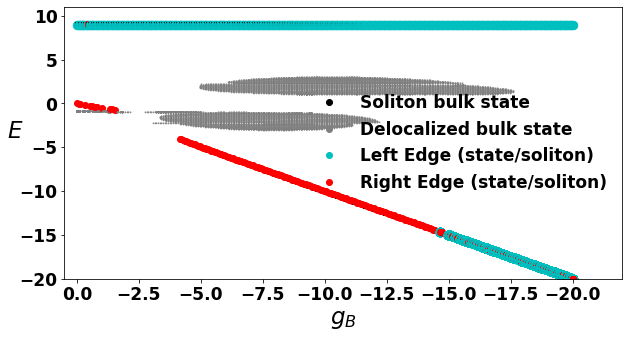}
    \caption{Energy vs parameter $g_{B}$ for $J_{1}=1~,~J_{2}=2~, ~v=0.0001~,~g_{A}=9~$ in the real space studies.}
    \label{Fig2.5}
\end{figure}

We now turn our attention to the case where $g_{A}$ and $g_{B}$ have a sign difference. There, it was found that delocalized bulk solutions persist in the vicinity of $|\frac{g_{A}}{g_{B}}|\simeq 1$, even at very high nonlinearity strengths as shown in Fig. \ref{Fig2.5}. This feature is not present in the case of uniform nonlinearity, i.e., $g_{A}=g_{B}$ \cite{2020-paper}, which solely supports solitons in the high nonlinearity regime. The persistence of the delocalized bulk solutions at very high nonlinearity is thus attributed by the neat interplay between the attractive and repulsive types of nonlinearities that coexist in the system. Indeed, the facts that the two delocalized bulk bands are centered around the values of $g_A$ and $g_B$ such that $|\frac{g_{A}}{g_{B}}|\simeq 1$ and that they terminate at large ratios of $|\frac{g_{A}}{g_{B}}|$ and $|\frac{g_{B}}{g_{A}}|$ further support this point.

An interesting observation in our real space studies is the identification of a state which is termed a ``Wave-Packet" (WP) state in the following. As the name suggests, the WP state is characterized by its localized but highly oscillatory feature that exists at sufficiently large nonlinearity strengths. Such a state was also identified in the uniform nonlinearity limit $g_A=g_B$, as presented in Fig.~7d of Ref.~\cite{2020-paper}, though it was not referred to as the WP state. Here, we shall examine such a WP state in more detail. In particular, Figures.~\ref{ch4,2} and ~\ref{ch4,2,5} depict the spatial profiles of some relevant state that develops into a WP state when either $g_A$ or $g_B$ is large enough, i.e., panel (c) in both figures. These figures further demonstrate that the WP state in fact originates from one of the delocalized bulk states; more specifically, it corresponds to the $k=0$ state in the linear limit. As elucidated in Ref.~\cite{2020-paper}, the localization of the WP state could be attributed to the interplay between topology and nonlinearity. That is, due to the state dependent Hamiltonian, the existence of a significant peak in a stationary state induces an effective localized potential at the location of the peak, which in turn serves as an effective edge that supports oscillatory-looking ``edge states" from either of its sides due to the topology of the underlying linear model. This argument also implies that such a WP state may exist both in the ``topologically trivial" and ``topologically nontrivial" regime, i.e., the regime $J_2<J_1$ and $J_1<J_2$ respectively, as the bulk of one regime is related to the other by a sublattice shift.  

\begin{figure}[!]
\begin{subfigure}{0.48\textwidth}
  \centering
  \includegraphics[width=1\linewidth]{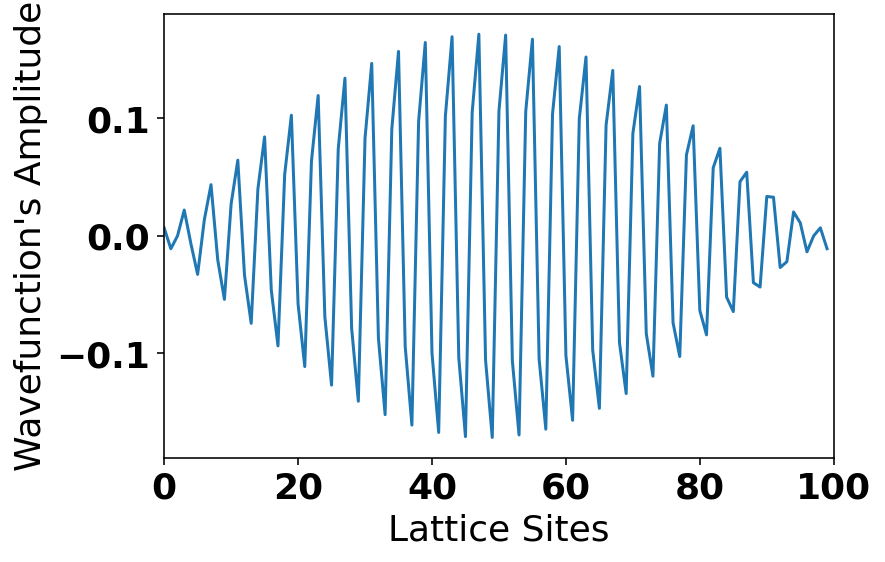}
  \caption{$g_{A}=0$ (linear case)}
  %\label{Fig2}
\end{subfigure}%

\begin{subfigure}{0.48\textwidth}
  \centering
  \includegraphics[width=1\linewidth]{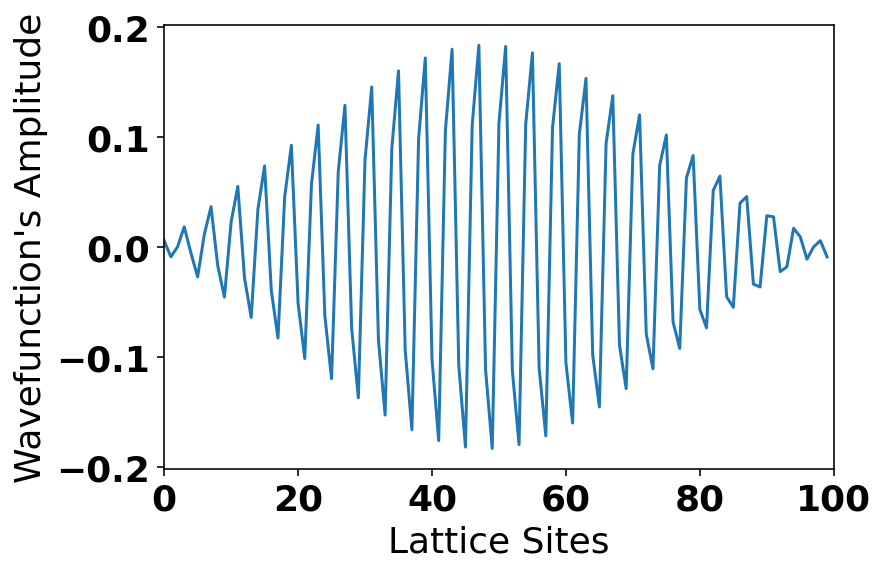}
  \caption{$g_{A}=2$}
  %\label{Fig2}
\end{subfigure}%

\begin{subfigure}{0.48\textwidth}
  \centering
  \includegraphics[width=1\linewidth]{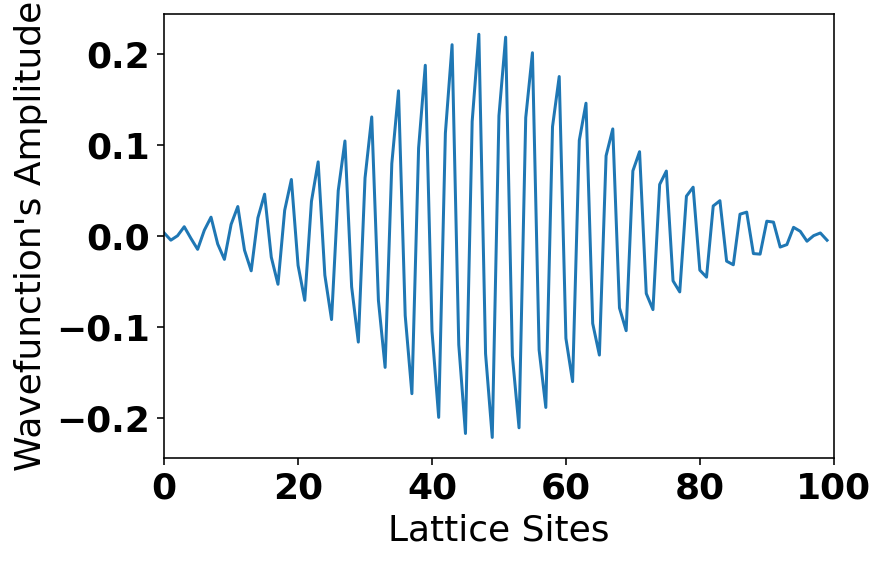}
  \caption{$g_{A}=6$}
  %\label{Fig2}
\end{subfigure}%
\caption{The development of the WP state as nonlinearity strength $g_{A}$ increases, starting from $g_{A}=0$ (top), to $g_{A}=6$ (bottom) for $J_{1}=1~,~J_{2}=2~, ~v=0.5~,~g_{B}=0~$. The typical profile of the WP state is illustrated in panel (c).}
\label{ch4,2}
\end{figure}

\begin{figure}[!]
\begin{subfigure}{0.48\textwidth}
  \centering
  \includegraphics[width=1 \linewidth]{FMSNoNonlinearity.png}
  \caption{$g_{B}=0$ (linear case)}
  %\label{fig:sfig1}
\end{subfigure}%

\begin{subfigure}{0.48\textwidth}
  \centering
  \includegraphics[width=1\linewidth]{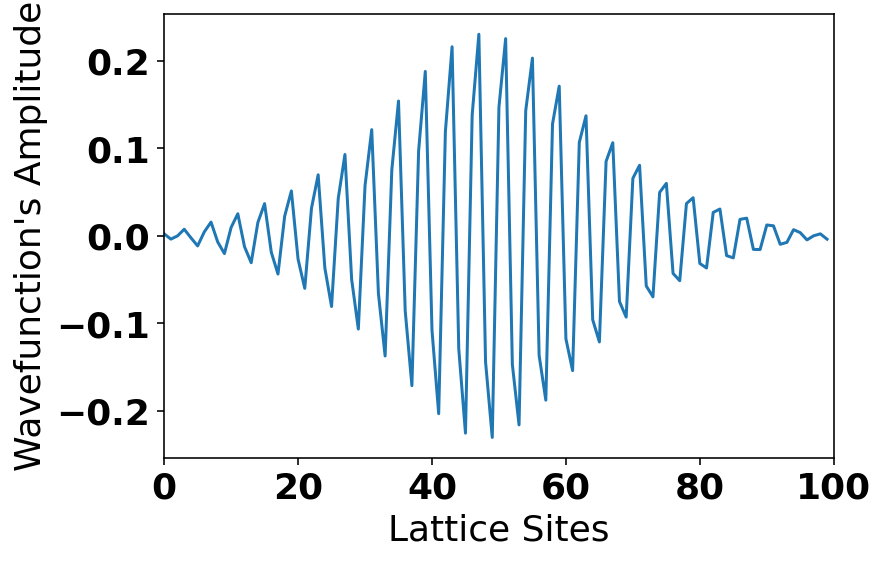}
  \caption{$g_{B}=1$}
  %\label{fig:sfig2}
\end{subfigure}

\begin{subfigure}{0.48\textwidth}
  \centering
  \includegraphics[width=1\linewidth]{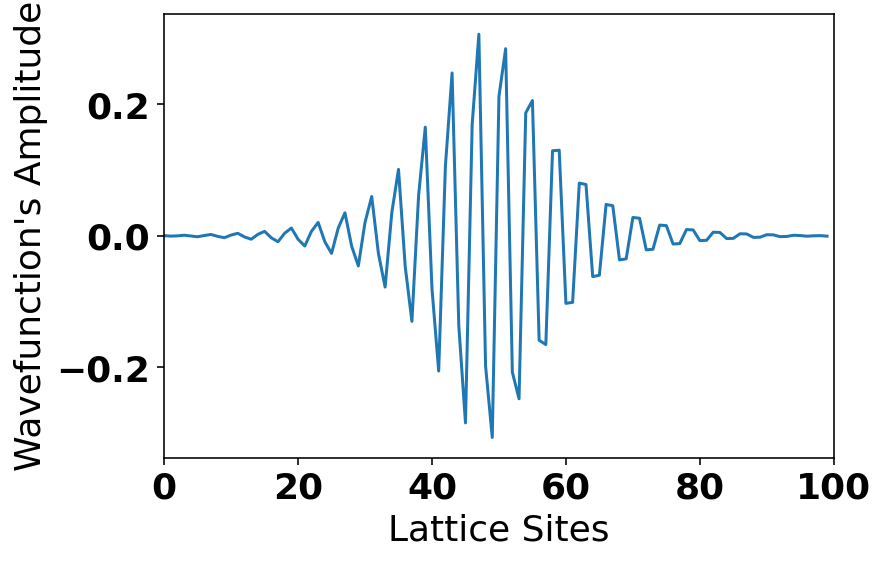}
  \caption{$g_{B}=2$}
  %\label{fig:sfig2}
\end{subfigure}
\caption{The development of the WP state as nonlinearity strength $g_{B}$ increases, starting from $g_{B}=0$ (top) to $g_{B}=2$ (bottom) for $J_{1}=1~,~J_{2}=2~, ~v=0.5~,~g_{A}=0~$.}
\label{ch4,2,5}
\end{figure}

In Figs.~\ref{ch4,1} and \ref{ch4,1.5}, a typical energy spectrum of our nonlinear system in the regime $J_1<J_2$ is presented as a function of $g_A$ and $g_B$ respectively. In Fig.~\ref{ch4,1}, the WP state corresponds to the persisting solution at the top of the bottom (delocalized) band, which occurs at energy that appears independent of $g_A$ and extends to very large $g_A$ values. While the WP state also emerges at varying $g_B$, i.e., Fig.~\ref{ch4,1.5}, it only extends to some moderate value of $g_B$ before it disappears, and its energy appears to slightly depend on $g_B$. The asymmetry between the behavior of the WP state at varying $g_A$ and $g_B$ could be attributed to the presence of the onsite potential $v$ that introduces some imbalance between the A and B sublattices. Indeed, choosing a smaller value of $v$, e.g., Fig.~\ref{Fig1}, reduces the maximum value of $g_A$ that supports the WP state. Finally, we have also verified through Fig.~\ref{ch4,3} that qualitatively the same behavior of the WP state is also observed in the regime $J_2<J_1$, as expected.

\begin{figure}
    \centering
    \includegraphics[width=1\linewidth]{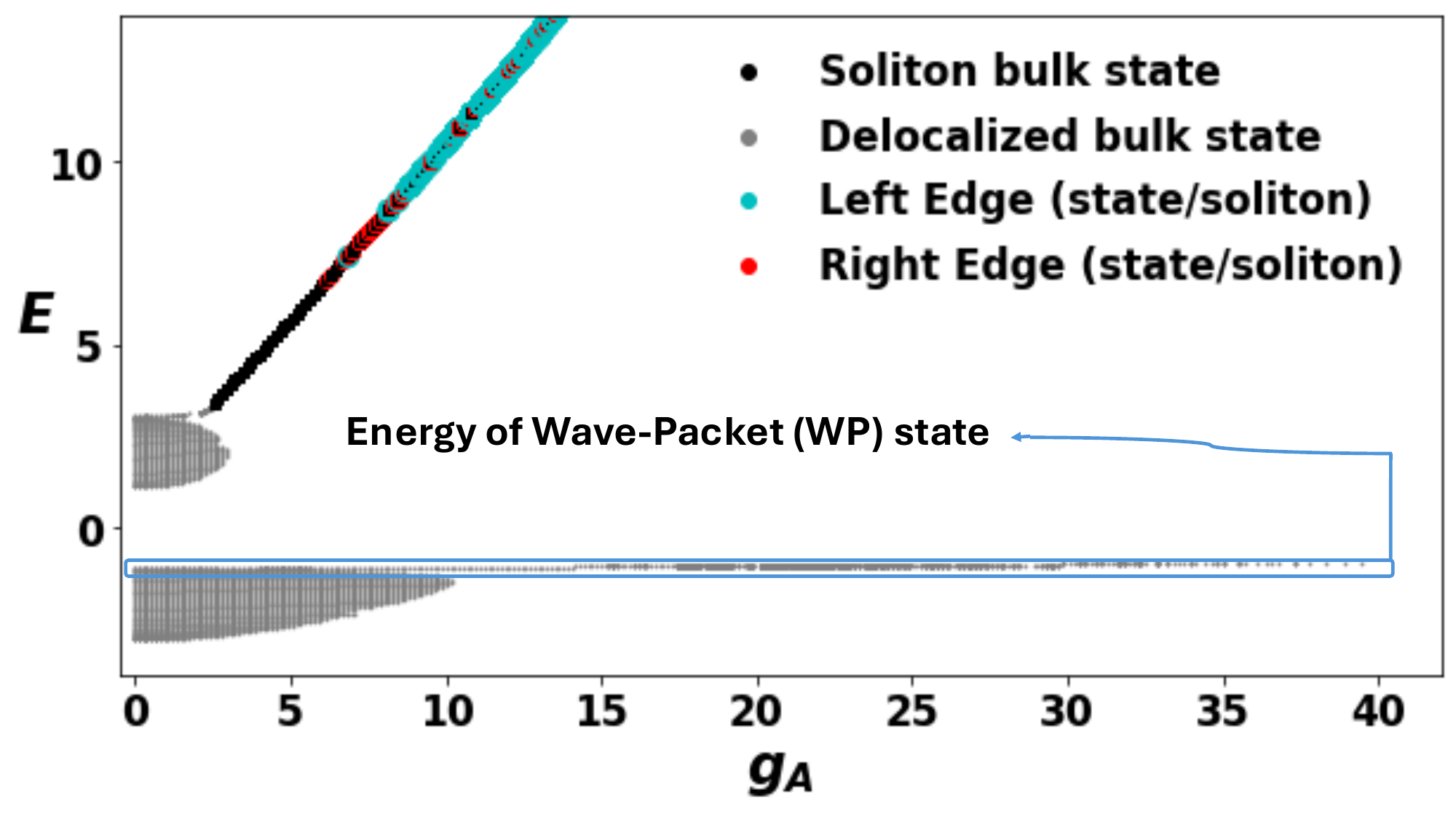}
    \caption{Energy vs parameter $g_{A}$ for $J_{1}=2~,~J_{2}=1~, ~v=0.5~,~g_{B}=0~$, i.e., the linear limit is topologically trivial, in the real space studies.}
    \label{ch4,3}
\end{figure}

\section{Concluding remarks}
\label{conc}

In this paper, we had considered a sublattice-dependent onsite nonlinearity in the topological SSH model and examined it both from a momentum space and a real space perspective. The momentum space analysis (with PBC) was first made by only considering Bloch's state solutions, so that semi-analytical approach could be made to obtain energy bands, perform dynamical stability analysis, and compute its Zak phases. In particular, the presence of nonlinearity induces some necessary modification to the expression of the Zak phase, which we carefully derived for a general system involving onsite nonlinearity, then numerically calculated it for our specific nonlinear system. In the case where $g_{A}$ and $g_{B}$ are positive and sufficiently large, energy gap-closing accompanied by discontinuity in the nonlinear Zak phase was observed, indicating a nonlinearity-induced topological phase transition.

The real space analysis (with OBC) was then made to obtain more general solutions, excluding the Bloch ones. In this case, a numerical approach based on the SCF Iterative Method was employed. For the iteration process, all energy eigenstates from the corresponding linear model were selected as initial trial states, and only sufficiently quickly converging nonlinear solutions were collected and plotted. In the case where $g_{A}$ and $g_{B}$ were both positive, it was found that in the topologically nontrivial regime ($J_1 < J_2$) and at sufficiently high nonlinearities, the energy of the left (right) edge state depends only on $g_{A}$ ($g_{B}$) and is constant with respect to $g_{B}$ ($g_{A}$). Moreover, the existence of only one edge state was found in some parameter regime. In addition, at sufficiently high nonlinearities, we found the presence of a touching point between two energy solutions, which only shift (and not disappear) under perturbation, thereby indicating a potential topological origin analogous to the Weyl semimetal. Finally, while all solutions are typically expected to be localized in general nonlinear systems at sufficiently high nonlinearity, we found that in our model, when $g_A$ and $g_B$ have opposite signs, some delocalized solutions were observed to persist even at extremely high nonlinear strengths. 

While two different approaches have been employed capturing OBC and PBC cases, the solutions presented in the paper are by no means complete. In particular, our PBC analysis is restricted only to finding Bloch solutions that are periodic by two sites. Generally, such a nonlinear system may also support Bloch-type solutions that are periodic by more than two-sites. This class of solutions are expected to exhibit interesting properties beyond those of the usual Bloch solutions, and are thus worth exploring in a separate study. Meanwhile, our OBC analysis is restricted only to finding converging solutions within our employed numerical method. A more careful investigation involving more sophisticated numerical methods, potentially in combination with some analytical treatment through appropriate approximations may form another interesting subject for further exploration of our model.

Incorporating other physical effects such as non-Hermiticity and/or time-periodic potential are expected to enrich the system even further is another potential avenue for future work. Indeed, similarly to how nonlinearity modifies some known properties of standard quantum mechanics, both non-Hermiticity and time periodicity also yield further deviations from the standard theory. For instance, non-Hermiticity removes the realness restriction of the energy eigenvalues, thereby enabling non-Hermitian systems to exhibit nontrivial topology at the energy eigenvalue level (point-gap topology) Refs. \cite{point-gap1, point-gap2, point-gap3}. Meanwhile, time-periodicity replaces the notion of energy, which is unbounded in nature, by quasienergy that is only defined modulo the driving frequency. The periodic nature of quasienergy may in turn facilitate additional topological phase transitions that otherwise cannot exist in undriven systems, resulting in a plethora of topological phenomena with no static counterparts Ref. \cite{time-periodicity, time-periodicity1}. In view of this, combining nonlinearity with any of these effects necessitates highly nontrivial analysis, but it comes with a promising prospect for discovering new physics.

Other potential directions for future studies involve considering the same type of onsite sublattice-dependent nonlinearity here in another lattice system such as the second-order topological insulator \cite{Benalcazar1, Benalcazar2, Raditya2019, ref1, ref2, ref3, ref4, ref5, ref6, ref7, ref8, ref9, ref10, ref11, ref12, ref13, ref14, ref15, ref16, ref17, ref18, ref19, ref20, ref21, ref22, ref23} or some extended SSH model \cite{nonlinearpot5, nonlinearpot6, nonlinearpot7, nonlinearpot8, nonlinearpot10, nonlinearpot11, nonlinearpot12, nonlinearpot13, ex-ssh1, ex-ssh4}. In this case, the resulting nonlinear second-order topological insulator serves as a natural extension of this work from one to two dimensions. Meanwhile, given that the extended SSH models of Refs. \cite{nonlinearpot5, nonlinearpot6, nonlinearpot7, nonlinearpot8, nonlinearpot10, nonlinearpot11, nonlinearpot12, nonlinearpot13, ex-ssh1, ex-ssh4} support non-zero energy edge states, adding nonlinearity is expected to yield richer phenomena that are not present in this work.  
 
\newpage

\appendix

\section{Self-Consistent Derivation of the Quartic Polynomial}
\label{Quartic}

{Recall that for a two-level Hamiltonian $H=H(\Sigma,k)$ of the form in Eq. (\ref{eq:rec3})
\begin{equation*}
H(\Sigma,k) = h_{x}(k) \sigma_{x} + h_{y}(k) \sigma_{y} + h_{z}(\Sigma) \sigma_{z} + \frac{S(\Sigma)}{2} I ,
\end{equation*} the characteristic polynomial reads:
\begin{equation}\label{characteristic}
{\rm det}\left[H-E(k) ~I\right]= E^{2} - S~E + \left( \frac{S}{2}\right)^{2} - \left( h_x^{2} + h_{y}^{2}+ h_{z}^{2}\right)= 0 .
\end{equation} To solve the state-dependent nonlinear eigenvalue problem presented in Eq. (\ref{eq:rec2}), one may without loss of generality, parametrize the state $\Phi$ (using Bloch sphere representation) as: 
\begin{equation}
 \Phi = 
\begin{pmatrix}
\varphi_{1}\\
\varphi_{2}
\end{pmatrix}  = 
\begin{pmatrix}
\cos\frac{\theta}{2}\\
e^{i\phi} \sin\frac{\theta}{2}
\end{pmatrix} .
\end{equation}
where $\theta \in [0, ~\pi]$, and $\phi \in [0,~2\pi)$. By squaring the geometric identity \begin{equation} \label{geometric}
h_{z}(\Sigma)=\sqrt{h_{x}^{2}+h_{y}^{2}+h_{z}^{2}}~  \cos{\theta} ~,
\end{equation} obtained from Bloch sphere representation, then inserting it into Eq. (\ref{characteristic}), and making use of the normalization condition $|\varphi_{1}|^2+|\varphi_{2}|^2=1$ to replace $|\varphi_{2}|^2$ in the resultant expression with $|\varphi_{2}|^2 = 1-|\varphi_{1}|^2= 1-\cos^{2}\frac{\theta}{2} $,
one gets a quartic polynomial in $|\varphi_{1}|^2=\cos^{2}\frac{\theta}{2}$ for a given $k$ value, which reads,
\begin{equation} \label{quartic_poly}
\small
\begin{split}
&x^{4} \left[(g_{A})^{2} + (g_{B})^{2} + 2g_{A}~g_{B} \right] + x^{3} [4~v~(g_{A}+g_{B}) - 4~g_{A}~g_{B}\\
&- (g_{A})^{2} -3~(g_{B})^{2}] + x^{2} [4~(J_{1}^{2} ~+ J_{2}^{2}~+v^{2}~+ 2~J_{1}~J_{2}~\cos{k})\\
&+3~g_{B}^{2} ~+ 2g_{A}~g_{B} - 4v~g_{A} -8v~g_{B}] + x~[ 4v~g_{B} -4~(J_{1}^{2} ~+ J_{2}^{2}~\\
&+v^{2}~+ 2~J_{1}~J_{2}~\cos{k}) -g_{B}^{2}] + J_{1}^{2} ~+ J_{2}^{2}~+ 2~J_{1}~J_{2}~\cos{k} = 0 ,
\end{split}
\end{equation}}where $x=\cos^{2}\frac{\theta}{2}$~, for notational convenience. The quartic polynomial Eq. (\ref{quartic_poly}) is then solved numerically for a given $k$ value, where $k \in [0,~2\pi)$, and only real roots $x \in \mathbb{R}$, with their values being  $0 \leq x \leq 1$, are selected. The latter condition follows from the normalization condition. Note that in the $x$ notation, $\Sigma = 1 - 2x$. It is now straight forward to evaluate corresponding energies from Eq.(\ref{eq:rec4}). Moreover, note that $\varphi_{1} = \sqrt{x}$ and $\varphi_{2}=\frac{h_{x}(k) + ih_{y}(k)}{\sqrt{h_{x}^{2}+ h_{y}^{2}}}\sqrt{1-x}$ where the expression in Eq. (\ref{eq:a6.5}) (from Appendix Sec.(\ref{Appen-Zak})) has been used. Evaluating $\varphi_{1}$ and $\varphi_{2}$ numerically is needed for numerical calculation of the general nonlinear Zak phase. 

\section{General Expression of Nonlinear Zak Phase}
\label{Appen-Zak}

{
In what follows, we derive a general expression of the nonlinear Zak phase, which holds for any well-behaved two-level real Hamiltonian with $C^{1}$ state-dependent (through $\Sigma= |\psi_{2}(t)|^2 - |\psi_{1}(t)|^2$) onsite nonlinearities. Consider a time-dependent nonlinear Schrodinger equation,
\begin{equation} \label{eq:a1}
i \frac{\partial \Psi(t)}{\partial t} = H(\Sigma, k) \Psi(t) ~,
\end{equation}
with $\hbar = 1$, $\Psi(t) = 
\begin{pmatrix}
\psi_{1} (t)\\
\psi_{2} (t)
\end{pmatrix} $, $\Sigma = |\psi_{2}(t)|^2 - |\psi_{1}(t)|^2 $ and the Hamiltonian has the form:
\begin{equation} \label{eq:a2}
H(\Sigma,k) = h_{x}(k) \sigma_{x} + h_{y}(k) \sigma_{y} + h_{z}(\Sigma) \sigma_{z} + \frac{S(\Sigma)}{2} I ~,
\end{equation} where $k$ is the quasimomentum,  $h_{x}(k)$ and $h_{y}(k)$ are well behaved real functions of $k$, $h_{z}(\Sigma) $ and $ S(\Sigma)$ are real $C^{1}$ functions of $ \Sigma $, $\sigma_{x}~,~ \sigma_{y}$ and $\sigma_{z}$ are the $2\times 2$ Pauli matrices, and $I$ is the $2\times 2$ identity matrix. Note that $ \Sigma $, being a state-dependent real variable, generally also depends implicitly on $k$. 

%\footnotetext{\textsuperscript{1} $h_{x}(k) , h_{y}(k)$ are to be fixed later to be the coupling terms in the SSH model}}

Considering $\Psi(t) = e^{i f(t)} \Phi$ where $f(t)$ is a real scalar function, and $\Phi = 
\begin{pmatrix}
\varphi_{1}\\
\varphi_{2}
\end{pmatrix} $. This form of $\Psi(t)$ was considered to separate explicit time-dependence in one term, namely $e^{i f(t)}$, note however that $\Phi$ have implicit time-dependence through an adiabatic parameter. Plugging $\Psi(t)$ into the nonlinear Schrodinger equation Eq. (\ref{eq:a1}), we get:
\begin{equation} \label{eq:a3}
\frac{\dd{f}}{\dd{t}} \Phi = i \frac{\dd{\Phi}}{\dd{t}} - H \Phi~.
\end{equation}
Multiplying Eq. (\ref{eq:a1}) from the left hand side by $\Psi(t)^{\dagger}$, we get the scalar equation:
\begin{equation} \label{eq:a4}
\frac{\dd{f}}{\dd{t}} = i \Phi^{\dagger}\frac{\dd{\Phi}}{\dd{t}} -\Phi^{\dagger}H \Phi ~.
\end{equation}
%\begin{equation}
%\left( \frac{\dd{f}}{\dd{t}} \right)\Phi = i \left(\Phi^{+}\frac{\dd{\Phi}}{\dd{t}}\right)\Phi - (\Phi^{+} H \Phi)\Phi
%\end{equation}
Now, we perturbatively expand the following in terms of an adiabatic parameter $\epsilon$:
\begin{equation} \label{eq:a5}
\begin{split}
\frac{\dd{f}}{\dd{t}}& = \alpha_{0} + \alpha_{1} \epsilon + \alpha_{2} \epsilon^{2} + \dots \\
\Phi& = \Phi^{(0)} + \epsilon \Phi^{(1)}+ \epsilon^{2} \Phi^{(2)}+ \dots \\
H& = H^{(0)} + \epsilon H^{(1)} + \epsilon^{2} H^{(2)} + \dots 
\end{split}
\end{equation} where $ H^{(0)} = \evalat{H(\Sigma, k)}{\epsilon=0} = H(\Sigma^{(0)}, k)$ and $H^{(1)}=\evalat{\frac{\dd{H}}{\dd{\epsilon}}}{\epsilon=0}=\evalat{\frac{\dd{\Sigma}}{\dd{\epsilon}}}{\epsilon=0} \Bigl[\evalat{\frac{\dd{h_{z}}}{\dd{\Sigma}}}{\Sigma=\Sigma^{(0)}}  \sigma_{z} + \frac{1}{2} \evalat{\frac{\dd{S}}{\dd{\Sigma}}}{\Sigma=\Sigma^{(0)}}  I\Bigr] $, with $\Sigma^{(0)}$ denoting $\Sigma$ evaluated at $\epsilon=0$. 

Next, we impose a condition (which states that the system is initially prepared in some stationary state $\Phi_{E}$):
\begin{equation} \label{eq:a6}
H^{(0)} \Phi_{E} = E \Phi_{E}~ ,
\end{equation}\\
 with $\Phi_{E} = \Phi^{(0)}$. Without loss of generality, we may parametrize an initial state $\Phi^{(0)}$ (using Bloch sphere representation) as: 
\begin{equation} \label{eq:a7}
 \Phi^{(0)} = 
\begin{pmatrix}
\varphi_{1}^{(0)}\\
\varphi_{2}^{(0)}
\end{pmatrix}  = 
\begin{pmatrix}
\cos\frac{\theta}{2}\\
e^{i\phi} \sin\frac{\theta}{2}
\end{pmatrix}~,
\end{equation}
Where $\theta \in [0, ~\pi]$ is the polar angle, and $\phi \in [0,~2\pi)$ is the azimuthal angle, with $e^{i \phi}$ being \cite{book}

\begin{equation} \label{eq:a6.5}
e^{i \phi} = \frac{h_{x}(k) + ih_{y}(k)}{\sqrt{h_{x}^{2}+ h_{y}^{2}}} .
\end{equation}
In addition, the normalization condition at the beginning of the adiabatic process (when the system is at $\Phi^{(0)}$) demands that
\begin{equation} \label{eq:a8}
|\varphi_{1}^{(0)}|^2 + |\varphi_{2}^{(0)}|^2 = 1 .
\end{equation} 
By further substituting $\Phi$ from Eq. (\ref{eq:a5}) to the above, i.e.,
\begin{equation*}
\begin{split}
|\psi_{1}(t)|^2 + |\psi_{2}(t)|^2 = |\varphi_{1}|^2 + |\varphi_{2}|^2 =1 ,
\end{split}
\end{equation*} 
and equating terms up to $\epsilon^{1}$, we obtain the condition:
\begin{equation} \label{eq:a9}
\begin{split}
\Re{\left(\varphi_{2}^{(0)\ast}\varphi_{2}^{(1)}\right)} = - \Re{\left(\varphi_{1}^{(0)\ast}\varphi_{1}^{(1)}\right)} \\
\Rightarrow \Re{\left ( e^{-i\phi} \varphi_{2}^{(1)}\right)} = -\cot{\left(\frac{\theta}{2}\right)} \Re{\left(\varphi_{1}^{(1)}\right)}
\end{split}
\end{equation} After substituting $\Phi$ from Eq. (\ref{eq:a5}) in $\Sigma$, and making use of Eqs. (\ref{eq:a7}, \ref{eq:a9}), $\Sigma$ has the form (up to $\epsilon^{1}$ terms):
\begin{equation}
\scalebox{1}{%
$
\begin{split}
\small
\Sigma &= |\psi_{2}(t)|^2 - |\psi_{1}(t)|^2 = |\varphi_{2}|^2 - |\varphi_{1}|^2 \\
&= |\varphi_{2}^{(0)}|^2 - |\varphi_{1}^{(0)}|^2 - 4 \epsilon \Re{ \left( \varphi_{1}^{(0)\ast}\varphi_{1}^{(1)}\right )} + \dots 
%= \Sigma^{(0)} - 4 \epsilon \Re{ \left( \varphi_{1}^{(0)\ast}\varphi_{1}^{(1)}\right )} + \dot 
\end{split}$}
\end{equation}
with $\Sigma^{(0)} = |\varphi_{2}^{(0)}|^2 - |\varphi_{1}^{(0)}|^2 = - \cos{\theta}$. 
Going back to Eq. (\ref{eq:a4}), and expanding it up to $\epsilon^{1}$, we get the two conditions
\begin{equation} \label{eq:a9.5}
\alpha_{0} = - E ~,
\end{equation}
\begin{equation} \label{eq:a10}
\epsilon \alpha_{1} = i \Phi^{(0)\dagger}\frac{\dd{\Phi^{(0)}}}{\dd{t}} - \epsilon \Phi^{(0)\dagger}H^{(1)} \Phi^{(0)}~.
\end{equation} where the eigenvalue equation $H^{(0)} \Phi^{(0)} = E \Phi^{(0)}$ was also used. 

Moreover, after substituting expanded $\frac{\dd{f}}{\dd{t}}~, \Phi$ and $H$ in Eq. (\ref{eq:a3}) and considering only the $\epsilon^{1}$ term of the first component's in the equation Eq. (\ref{eq:a3}), we get

\begin{equation*} 
\epsilon\alpha_{1}\varphi_{1}^{(0)} = i \frac{\dd{\varphi_{1}^{(0)}}}{\dd{t}} - \epsilon \Bigl[\alpha_{0}~\varphi_{1}^{(1)} + H_{11}^{(0)}\varphi_{1}^{(1)} + H_{12}^{(0)}\varphi_{2}^{(1)}+ H_{11}^{(1)}\varphi_{1}^{(0)}\Bigr]
\end{equation*}
\begin{equation} \label{eq:a11}
\scalebox{0.8}{%
$
\begin{split}
\small
\epsilon\alpha_{1}\varphi_{1}^{(0)} = i \frac{\dd{\varphi_{1}^{(0)}}}{\dd{t}} - \epsilon \Bigl[\alpha_{0}~\varphi_{1}^{(1)} + \left(\evalat{h_{z}(\Sigma)}{\epsilon=0} + \frac{\evalat{S(\Sigma)}{\epsilon=0}}{2} \right)\varphi_{1}^{(1)} \\
+ \left( h_{x}(k) - ih_{y}(k)\right) \varphi_{2}^{(1)} + ~\evalat{\frac{\dd{\Sigma}}{\dd{\epsilon}}}{\epsilon=0} \left( \evalat{\frac{\dd{h_{z}}}{\dd{\Sigma}}}{\Sigma=\Sigma^{(0)}} + \frac{1}{2} \evalat{\frac{\dd{S}}{\dd{\Sigma}}}{\Sigma=\Sigma^{(0)}}\right) \varphi_{1}^{(0)}\Bigr]
\end{split}$}
\end{equation}
Taking the real component of both sides of Eq. (\ref{eq:a11}), we get,
\begin{equation} \label{eq:a12}
\scalebox{0.8}{%
$
\begin{split}
\small
\epsilon\alpha_{1}\varphi_{1}^{(0)} = i \frac{\dd{\varphi_{1}^{(0)}}}{\dd{t}} - \epsilon \Bigl[\alpha_{0}~ \Re{(\varphi_{1}^{(1)})} + \left(\evalat{h_{z}(\Sigma)}{\epsilon=0} + \frac{\evalat{S(\Sigma)}{\epsilon=0}}{2} \right)\Re{(\varphi_{1}^{(1)})} \\
+ \Re{\left[\left( h_{x}(k) - ih_{y}(k)\right) \varphi_{2}^{(1)}\right]} + ~\evalat{\frac{\dd{\Sigma}}{\dd{\epsilon}}}{\epsilon=0} \left( \evalat{\frac{\dd{h_{z}}}{\dd{\Sigma}}}{\Sigma=\Sigma^{(0)}} + \frac{1}{2} \evalat{\frac{\dd{S}}{\dd{\Sigma}}}{\Sigma=\Sigma^{(0)}}\right) \varphi_{1}^{(0)}\Bigr]
\end{split}$}
\end{equation} Making use of Eqs. (\ref{eq:a6.5}, \ref{eq:a9}), we replace $\Re{\left[\left( h_{x}(k) - ih_{y}(k)\right) \varphi_{2}^{(1)}\right]}$ in Eq. (\ref{eq:a12}) with $-\sqrt{h_{x}^{2}+ h_{y}^{2}}~ \cot{\left(\frac{\theta}{2}\right)}\Re{\left(\varphi_{1}^{(1)}\right)}$. After using Eq. (\ref{eq:a9.5}), then evaluating $\evalat{\frac{\dd{\Sigma}}{\dd{\epsilon}}}{\epsilon=0} = -4~\varphi_{1}^{(0)} \Re{(\varphi_{1}^{(1)})}$ and recalling that $\varphi_{1}^{(0)}$ is real in our case, we get,
\begin{equation*}
\scalebox{0.8}{%
$
\begin{split}
\epsilon \alpha_{1} \varphi_{1}^{(0)} = i \frac{\dd{\varphi_{1}^{(0)}}}{\dd{t}} - \epsilon~\Re{(\varphi_{1}^{(1)})} \Big\{ -E + \evalat{h_{z}(\Sigma)}{\epsilon=0} 
+ \frac{\evalat{S(\Sigma)}{\epsilon=0}}{2} \\  - \sqrt{h_{x}^{2} + h_{y}^{2}} \cot\frac{\theta}{2} - 4 \left(\varphi_{1}^{(0)}\right)^{2} \Bigl[\evalat{\frac{\dd{h_{z}}}{\dd{\Sigma}}}{\Sigma=\Sigma^{(0)}}  + \frac{1}{2} \evalat{\frac{\dd{S}}{\dd{\Sigma}}}{\Sigma=\Sigma^{(0)}} \Bigr] \Big\}
\end{split}$}
\end{equation*}
Substituting by $\varphi_{1}^{(0)}= \cos{\frac{\theta}{2}}$ then solving for $\epsilon~ \Re{(\varphi_{1}^{(1)})}$, we get,
\begin{widetext}
\begin{equation}\label{eq:a13}
\scalebox{1}{%
$\epsilon\Re{(\varphi_{1}^{(1)})} = \frac{i \frac{\dd{\varphi_{1}^{(0)}}}{\dd{t}} - \epsilon \alpha_{1} \cos{\frac{\theta}{2}}}{\Big\{-E + \evalat{h_{z}(\Sigma)}{\epsilon=0} + \frac{1}{2}\evalat{S(\Sigma)}{\epsilon=0}  - \sqrt{h_{x}^{2} + h_{y}^{2}} \cot\frac{\theta}{2} - 4 \cos^{2}\left(\frac{\theta}{2}\right) \Bigl[\evalat{\frac{\dd{h_{z}}}{\dd{\Sigma}}}{\Sigma=\Sigma^{(0)}}  + \frac{1}{2} \evalat{\frac{\dd{S}}{\dd{\Sigma}}}{\Sigma=\Sigma^{(0)}}\Bigr]\Big\}}$}~.
\end{equation} Recall that when using the explicit form of $H^{(1)}$, Eq. (\ref{eq:a10}) takes the form,
\begin{equation}\label{eq:a14}
\epsilon \alpha_{1} = i \Phi^{(0)\dagger}\frac{\dd{\Phi^{(0)}}}{\dd{t}} - 4\cos{\frac{\theta}{2}}\left(\evalat{\frac{\dd{h_{z}}}{\dd{\Sigma}}}{\Sigma=\Sigma^{(0)}} \Sigma^{(0)}-\frac{1}{2} \evalat{\frac{\dd{S}}{\dd{\Sigma}}}{\Sigma=\Sigma^{(0)}} \right)\epsilon \Re{(\varphi_{1}^{(1)})}~.
\end{equation} By substituting $\epsilon~\Re{(\varphi_{1}^{(1)})}$ from Eq. (\ref{eq:a13}) into Eq. (\ref{eq:a14}) then solving for $\epsilon \alpha_{1}$, we find,
\begin{equation}\label{eq:a15}
\epsilon \alpha_{1} = \frac{i \Phi^{(0)\dagger}\frac{\dd{\Phi^{(0)}}}{\dd{t}} + 4 i \cos\frac{\theta}{2} \frac{\dd{\varphi_{1}^{(0)}}}{\dd{t}} \left( \frac{\frac{1}{2} \evalat{\frac{\dd{S}}{\dd{\Sigma}}}{\Sigma=\Sigma^{(0)}} - \Sigma^{(0)} \evalat{\frac{\dd{h_{z}}}{\dd{\Sigma}}}{\Sigma=\Sigma^{(0)}}}{-E + \evalat{h_{z}(\Sigma)}{\epsilon=0} + \frac{1}{2}\evalat{S(\Sigma)}{\epsilon=0}  - \sqrt{h_{x}^{2} + h_{y}^{2}} \cot\frac{\theta}{2} - 4 \cos^{2}\left(\frac{\theta}{2}\right) \Bigl[\evalat{\frac{\dd{h_{z}}}{\dd{\Sigma}}}{\Sigma=\Sigma^{(0)}}  + \frac{1}{2} \evalat{\frac{\dd{S}}{\dd{\Sigma}}}{\Sigma=\Sigma^{(0)}}\Bigr]} \right)}{1 + 4 \cos^{2}\left(\frac{\theta}{2}\right) \left( \frac{\frac{1}{2} \evalat{\frac{\dd{S}}{\dd{\Sigma}}}{\Sigma=\Sigma^{(0)}} - \Sigma^{(0)} \evalat{\frac{\dd{h_{z}}}{\dd{\Sigma}}}{\Sigma=\Sigma^{(0)}}}{-E + \evalat{h_{z}(\Sigma)}{\epsilon=0} + \frac{1}{2}\evalat{S(\Sigma)}{\epsilon=0}  - \sqrt{h_{x}^{2} + h_{y}^{2}} \cot\frac{\theta}{2} - 4 \cos^{2}\left(\frac{\theta}{2}\right) \Bigl[\evalat{\frac{\dd{h_{z}}}{\dd{\Sigma}}}{\Sigma=\Sigma^{(0)}}  + \frac{1}{2} \evalat{\frac{\dd{S}}{\dd{\Sigma}}}{\Sigma=\Sigma^{(0)}}\Bigr]} \right)}~.
\end{equation}
\end{widetext}
which is a general expression for the nonlinear Zak phase. Please note that the expression holds for any real-valued well-behaved functions $h_{x}(k) , h_{y}(k)$, and for any real $C^{1}$ functions $h_{z}(\Sigma)$,~and $ S(\Sigma)$. Note also that the fraction within  $( ~~~) $ vanishes for linear Hamiltonians and the expression reduces to the conventional Zak phase $i \Phi^{(0)\dagger}\frac{\dd{\Phi^{(0)}}}{\dd{t}}$, as expected.  

Recall that the expression of the nonlinear Zak phase used in Ref.~\cite{2020-paper} reads,

\begin{equation} \label{ch4,4.2,0}
\epsilon\alpha_{1} = i \Phi^{\dagger}\frac{\dd{\Phi}}{\dd{t}}\left(1+\frac{\evalat{\frac{\dd{h_{z}}}{\dd{\Sigma}}}{\Sigma=\Sigma^{(0)}}\cos{\theta}(1+\cos{\theta})}{E+\evalat{\frac{\dd{h_{z}}}{\dd{\Sigma}}}{\Sigma=\Sigma^{(0)}}\sin^{2}{\theta}} \right)~.
\end{equation} 
Despite the algebraic difference between our expression of the general nonlinear Zak phase Eq. (\ref{eq:rec7}) in the limit $g_{A} = g_{B}$ and Eq. (\ref{ch4,4.2,0}), our numerical calculation shown in Fig. \ref{Zak-Phases-Reproduction} of the main text clearly demonstrates that the two yield the same result. The difference in the two expressions' algebraic forms might be attributed to how each expression was derived. In Ref.~\cite{2020-paper}, the derivation relied on using a state orthonormal to the unperturbed stationary state (which was termed ``the hidden eigenstate" in Appendix A of Ref.~\cite{2020-paper}), whereas in our derivation, we used Eq. (\ref{eq:a6.5}) instead. Therefore, while the algebraic equivalence between the two expressions may not be clearly established just by direct inspection, the facts that both expressions were derived self-consistently, yield the same result in numerical studies, and reduce to the expected result in the linear limit, strongly suggest their equivalence.

\section{Theory of Dynamical Stability Analysis}
\label{ch2,3}
{
In this section, we derive the analytical tools needed to apply a dynamical stability analysis on the stationary states associated with the energy bands in the Bloch solution studies of our nonlinear system. Consider a time-dependent Schrodinger equation and its complex conjugate equation,
\begin{equation} \label{eq:ab1}
\begin{split}
i \frac{\partial \Psi}{\partial t}& = H(\Sigma, k) \Psi~, \\
-i \frac{\partial \Psi^{\ast}}{\partial t}& = H(\Sigma, k)\Psi^{\ast} ~.
\end{split}
\end{equation}where $\hbar = 1$, $\Psi = 
\begin{pmatrix}
\psi_{1} \\
\psi_{2}
\end{pmatrix} $, $\Sigma = |\psi_{2}|^2 - |\psi_{1}|^2 $, $k$ is the quasimomentum, and the Hamiltonian $H=H(\Sigma, k)$ is assumed to be real (i.e. $H^{\ast}=H$).

Suppose the system is prepared in a state close to some stationary state $\Psi'\equiv \Psi^{(0)} +\delta \Psi$, where 
\begin{equation}\label{eq:ab3}
i \evalat{\frac{\dd{\Psi}}{\dd{t}}}{t=t_{0}} = H^{(0)}\Psi^{(0)}= E~\Psi^{(0)} ,
\end{equation}
$H^{(0)}= H(\Sigma^{(0)}, k)$, $\Psi^{(0)} = 
\begin{pmatrix}
\psi_{1}^{(0)} \\
\psi_{2}^{(0)}
\end{pmatrix} $, $\Sigma^{(0)} = |\psi_{2}^{(0)}|^2 - |\psi_{1}^{(0)}|^2$, and $\delta \Psi = 
\begin{pmatrix}
\delta \psi_{1} \\
\delta \psi_{2}
\end{pmatrix} $ is a small perturbation to $\Psi^{(0)}$. It then follows that, up to first order in the perturbation parameters,
\begin{eqnarray}\label{eq:ab4}
\Sigma^{\prime} &=& \Sigma^{(0)} + \psi_{2}^{\ast (0)}~\delta\psi_{2}+ \psi_{2}^{(0)}~\delta\psi_{2}^{\ast} - \psi_{1}^{\ast (0)}~\delta\psi_{1} \nonumber \\
&& - \psi_{1}^{(0)}~\delta\psi_{1}^{\ast} + \delta\psi_{2}^{\ast}~\delta\psi_{2} - \delta\psi_{1}^{\ast}~\delta\psi_{1} .
\end{eqnarray}
By plugging in Eq.~(\ref{eq:ab4}) to Eq.~(\ref{eq:ab1}), and further solving the system of equations in (\ref{eq:ab1}) for each $\delta\psi_{1} , ~\delta\psi_{2}, ~ \delta\psi_{1}^{\ast}$ and $\delta\psi_{2}^{\ast}$ independently, we arrive at
\begin{gather} \label{eq:ab6}
 i~\partial_{t}\begin{pmatrix}
\delta \psi_{1} \\
\delta \psi_{2} \\
\delta\psi_{1}^{\ast} \\
\delta\psi_{2}^{\ast}
\end{pmatrix}
 = \mathcal{L} \begin{pmatrix}
\delta \psi_{1} \\
\delta \psi_{2} \\
\delta\psi_{1}^{\ast} \\
\delta\psi_{2}^{\ast} 
\end{pmatrix} ,
\end{gather} where $\mathcal{L}$ is a $4\times4$ matrix that is generally non-hermitian and depends on the Hamiltonian in question. 

Equation~(\ref{eq:ab6}) implies that the time evolution of each perturbation component $\delta\psi_{1} , ~\delta\psi_{2}, ~ \delta\psi_{1}^{\ast}$ and $\delta\psi_{2}^{\ast}$ is governed by the eigenvalues $\{ \lambda_{n}, \forall n\}$ of $\mathcal{L}$ through $e^{-i~\lambda_{n}t}~$. As $\mathcal{L}$ is generally non-hermitian, its eigenvalues are generally complex. Accordingly, for the state's perturbation to not grow to infinity as time increases, the eigenvalues have to fulfill the following condition:
\begin{equation}\label{eq:ab7}
\Im{(\lambda_{n})} = 0 ~, ~ \forall n ,
\end{equation}
i.e., all eigenvalues have to be real. In this case, the corresponding state is then classified as dynamically stable. In our numerical calculations presented in the main text, we examine the quantity:
\begin{equation}\label{eq:ab8}
\max{[|\Im{(\lambda_{n})}|]} ~, ~ \forall n .
\end{equation}
Equivalently to the condition in Eq.~(\ref{eq:ab7}), if $\max{[|\Im{(\lambda_{n})}|]}=0~, ~ \forall n$, the state is dynamically stable. Otherwise, it is dynamically unstable. 
}

\label{app:0}

\clearpage
%\section{References}
%\label{References}

\end{document}